\documentclass[%
 reprint,
 amsmath,amssymb,
prstab,
floatfix,
]{revtex4-1}

\usepackage{subfig}
\usepackage{graphicx}
\usepackage{epstopdf}
\usepackage{dcolumn}
\usepackage{bm}

\newcommand{\figref}[1]{FIG.~\ref{#1}}
\newcommand{\tabref}[1]{TABLE ~\ref{#1}}

\newcommand{\Eqref}[1]{Equation \hspace{-0.2cm}~\eqref{#1}}

\begin{document}

\preprint{APS/123-QED}

\title{Evolution of a beam dynamics model for the transport line \\ in a proton therapy facility}

\author{V.\ Rizzoglio} 
\email{valeria.rizzoglio@psi.ch} 

\author{A.\ Adelmann}
\email{andreas.adelmann@psi.ch}

\author{C.\ Baumgarten} \author{M.\ Frey} \author{A.\ Gerbershagen} \email{Presently at CERN, 1211 Geneva, Switzerland}
\author{D.\ Meer} \author{J.\ M.\ Schippers}

\affiliation{Paul Scherrer Institut, 5232 Villigen, Switzerland}

\begin{abstract}
During the conceptual design of an accelerator or beamline, first-order beam dynamics models are essential for studying beam properties.\ However, they can only produce approximate results.\ During commissioning, these approximate results are compared to measurements, which will rarely coincide if the model does not include the relevant physics.\ It is therefore essential that this linear model is extended to include higher-order effects.\
In this paper, the effects of particle-matter interaction have been included in the model of the transport lines in the proton therapy facility at the Paul Scherrer Institut (PSI) in Switzerland.\ The first-order models of these beamlines provide an approximated estimation of beam size, energy loss and transmission.\ To improve the performance of the facility, a more precise model was required and has been developed with OPAL (Object oriented Particle Accelerator Library), a multi-particle open source beam dynamics code.\ 
In OPAL, the Monte Carlo simulations of Coulomb scattering and energy loss are performed seamless with the particle tracking.\ Beside the linear optics, the influence of the passive elements (e.g.\ degrader, collimators, scattering foils and air gaps) on the beam emittance and energy spread can be analysed in the new model.\ This allows for a significantly improved precision in the prediction of beam transmission and beam properties.\ The accuracy of the OPAL model has been confirmed by numerous measurements.
\end{abstract}

\maketitle


\section{Introduction}
\label{sec:intro}

Today several codes for beam dynamics and transport line simulations are available.\ Single- or multi- particle codes are used in a variety of applications.\ Evidently it is hard to cover all the physics processes with a single code, therefore, several codes have to be combined together to get an adequate model of a certain machine.

The preliminary lattice of an accelerator or beamline is normally designed and optimised with the beam dynamics models in linear approximation.\ With the evolution of the design and especially during the commissioning, a more accurate and precise beam dynamics model is required.\ The simplified linear model has to be extended to cover higher-order effects and, depending on the application, has to include collective effects or particle-matter interaction.\ In many cases this requires a combination of Monte Carlo simulations and particle tracking codes \cite{VanGoethem2009, Pavlovic2008}.

In this paper, we show the importance to include the effects of particle-matter interaction in the beam dynamics model of the transport lines in a proton therapy facility.

In particle therapy facilities, one of the main issues is the delivery of different beam energies to scan the tumor in depth.\ In a cyclotron-based facility this process is performed by means of the multiple scattering within a degrader.\ As a side effect of the degradation process the beam emittance and energy spread are increased \cite{Farley2005, Anferov2003}.\ To provide the required beam quality at the patient location (isocenter), the beam has to be shaped by a set of collimators and the energy spread is reduced by means of an Energy Selection System (ESS).\ Afterward, the beam crosses air gaps and several thin foils (inside beam profile and current monitors) that result in further scattering and emittance increase \cite{Paganetti2011}.

The complete characterisation of the beam quality along such beamlines requires the use of two different types of codes:\ beam dynamics codes for the optics simulations (e.g.\ TRANSPORT \cite{Transport}, TURTLE \cite{Turtle}, MAD-X \cite{Madx}) and Monte Carlo codes (e.g.\ GEANT4 \cite{Geant}, FLUKA \cite{Fluka}, TOPAS \cite{Topas}) for energy loss and scattering evaluation in the degrader, collimators \cite{VanGoethem2009} and nozzle (i.e.\ the last part of the beamline before the patient) \cite{Paganetti2004}.

The use of several different codes however is arduous, error-prone and time consuming.\ A single code that integrates particle tracking and Monte Carlo capabilities is therefore desired.

For more than 20 years PROSCAN, a cyclotron-based proton therapy facility, has been treating patients at the Paul Scherrer Institut (PSI) in Switzerland.\ In this paper, we discuss the limitations of the linear model of the transport lines in this facility and the improvement arising from a single beam dynamics code that includes also the particle-matter interaction.

In this facility, the optics of the beam transport lines was originally modelled with TRANSPORT and optimised experimentally \cite{Rohrer2002}.\ In 2014, the PROSCAN facility was extended with a third gantry.\ The purpose of the transport line toward the new gantry is to provide an energy-dependent intensity at the patient location.\ Therefore, a more precise evaluation of the beam losses, emittance and energy changes along the beamline is needed.\ Since TRANSPORT is not suitable for this kind of analysis, the multi-particle open source OPAL \cite{OPAL} framework has been used.\ OPAL is able to combine multi-particle tracking in linear and nonlinear regime with a Monte Carlo simulation of particle-matter interaction.\ The influence of the passive elements (e.g.\ degrader, collimators, scattering foils and air gaps) on the phase space of the beam can be analysed in detail.\ This leads to an enhanced accuracy in the evaluation of transmission, emittance and energy changes along the beamline.

To simplify the model setup and post-processing analysis, a versatile tool called ROGER (ROot GEnerator for Runopal) has been developed starting from the existing H5root framework \cite{H5root}.\ ROGER allows the access to the beamline setting, a direct comparison between the model and the measurements and a post-processing analysis of the model results.

In section \ref{sec:2} the facility layout and the TRANSPORT model are described.\ Section \ref{sec:3} is dedicated to the OPAL framework, to the implementation of the Monte Carlo model and to a short introduction to ROGER. In the last section the OPAL model for the beamline toward the new gantry is presented and benchmarked against different types of measurements. 


\section{PROSCAN facility and the TRANSPORT model}
\label{sec:2}

Tumor treatment with active scanning proton beams has been performed at PSI since 1996 and started with the Gantry 1 \cite{Pedroni1995}.

\begin{figure}[h!]
	\includegraphics[scale=0.1, keepaspectratio=true]{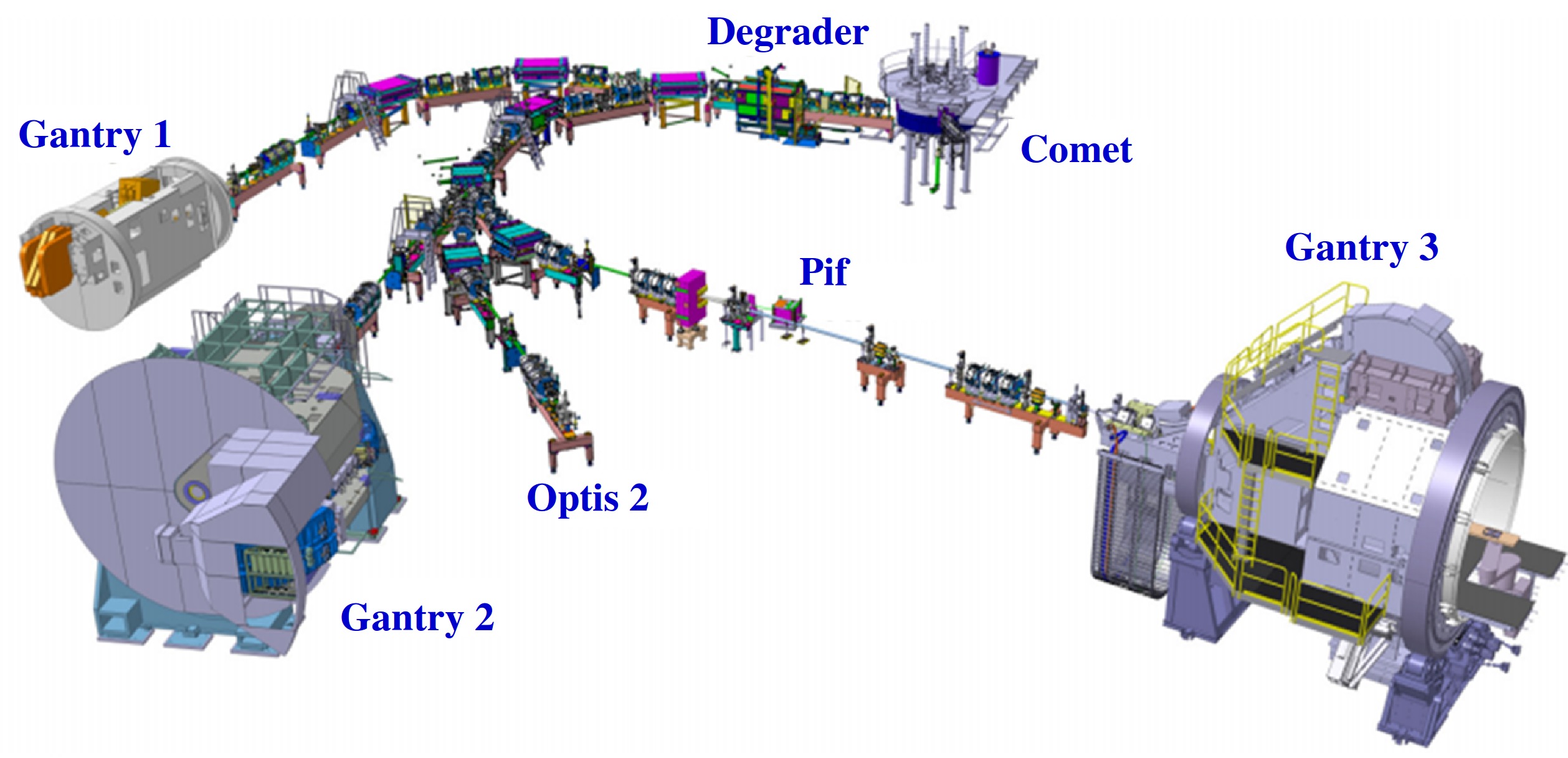}
	\caption{Schematic layout of the PROSCAN facility \cite{Koschik2015}.}
	\label{fig:Proscan}
\end{figure}

Today the facility consists of four clinical treatment rooms and one experimental area as shown in \figref{fig:Proscan}.\ In addition to Gantry 1, since 2013 the new Gantry 2 has been used for patient treatment with a fast spot scanning and optional fast rescanning technique \cite{Pedroni2004, Pedroni2011}.\ Based on the passive scattering technique, Optis 2 is a facility dedicated to eye-tumor treatment and consists of a horizontal fixed beam line \cite{optis}.\ Since 2014, the PROSCAN facility is expanded with Gantry 3, a new $360^{\circ}$ gantry built in a research collaboration between Varian Medical Systems \cite{Varian} and PSI.\ Finally the proton irradiation facility (Pif) provides the possibility for non-clinical irradiation experiments \cite{PIF}. 

Since 2007 the facility is equipped with a dedicated superconducting cyclotron Comet.\ A 250 MeV proton beam is extracted from Comet with a maximum current for the therapy up to 800 nA \cite{Schippers2007, Baumgarten2007}.\ The extracted beam is focused by a quadrupole triplet onto a degrader, which consists of 2 pairs of 3 movable graphite wedges (see section \ref{subsec:degrader} 
for more details).\ The amount of graphite that the beam has to pass is controlled by the position of the wedges.\ This allows the delivery of any proton energy in the range of 230$-$70 MeV.\ Behind the degrader, three collimators and the ESS are used to reduce emittance and momentum spread of the degraded beam.\ A geometric un-normalized emittance of typically \ 30 $\pi$ mm mrad and $\pm 1.0\%$ of momentum spread are required to match the beam within the acceptance of beamline and gantries.\ The proton beam is then guided by the transport lines to the selected treatment room.

\begin{figure*}[htb]
	\centering
	\includegraphics[scale=0.7, keepaspectratio=true]{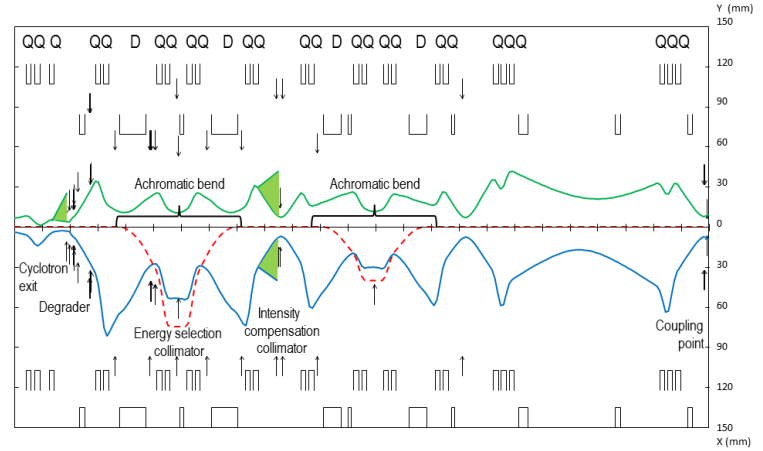}
	\caption{Beam envelope of the TRANSPORT model from the cyclotron exit up to the coupling point of the Gantry 3 \cite{Koschik2015}.\ The blue and the green lines represent the beam size in the horizontal-(bending) plane and in the vertical plane, respectively.\ The trajectory of a particle with 1\% momentum offset is marked by the red dashed line.\ The light-green areas indicate where the intensity compensation is applied.}
	\label{fig:Transport}
\end{figure*}

The beam optics of the five beamlines satisfies certain constraints:\ a symmetric double waist with dispersion-suppression at the coupling point of the gantries, achromatic bending sections and monotonic variation of the quadrupole currents with the beam rigidity to avoid hysteresis effects.\ Along the beamlines nearly 1:1 imaging conditions are established to have a better control on the beam size.\ A first-order optics model that satisfies these requirements was developed with the beam-envelope code TRANSPORT.\ Its fitting routine is well-suited to quickly provide a first approximate solution that matches the requirements.\ For the beamline toward Gantry 3, the TRANSPORT model is shown in \figref{fig:Transport}.

In addition to the energy reduction, in a proton therapy facility, another issue is to provide an adequate beam intensity at the isocenter.\ The beam intensity depends on the extraction efficiency of the accelerator and on the transmission through the following beamline.\ The ideal condition would be to maintain a constant beam intensity over the entire energy range (230$-$70 MeV).

At the PROSCAN facility, the beam emittance and energy spread increase in the degrader and the subsequent beam collimation and energy selection leads to a correspondingly strong and energy dependent decrease in beam intensity.\ The transmission between the maximum (230 MeV) and minimum (70 MeV) energy varies by a factor $10^3$.\ For reasons of patient safety and precision, the beam intensity should vary over the energy range by a factor $\leq$ 10 \cite{Baumgarten2015-2}.\ Therefore, this excess in the intensity variation over the energy range has to be compensated.\ This leads to an energy-dependent intensity modulation, also referred to as intensity compensation \cite{Koschik2015}.\ At the present time, two strategies have been developed for the intensity compensation: using the vertical deflector placed in the central region of the cyclotron Comet that provides a fast modulation of the extracted current \cite{Schippers2007-1} or dumping a fraction of the beam at specific locations along the beamlines.\ Based on the second strategy, the first-order beam optics of transport line toward Gantry 3 was optimized with TRANSPORT, providing the starting model for this work.   

Within the intensity compensation scheme, the beamline optics was developed with two different strategies depending on the energy range:

\vspace{-0.25cm}

\begin{itemize}
\item 70$-$140 MeV: maximised transmission by scaling the magnet currents with the beam rigidity
\vspace{-0.2cm}
\item 140$-$230 MeV: limited or reduced transmission by defocusing the beam at two specific locations: before the degrader and in front of the dedicated Intensity Compensation Collimator (ICC) after the ESS (see the light-green areas in \figref{fig:Transport}). 
\end{itemize}

\vspace{-0.25cm}

With this approach, a maximised beam intensity is expected at the isocenter for lower energies, while a reduced beam intensity for higher energies is foreseen \cite{Koschik2015, Baumgarten2015}. 

In the context of the intensity compensation, the partial cuts at collimators and the scattering of the beam has to be modelled properly in order to obtain an accurate evaluation of the total transmission.\ However, TRANSPORT is not well-suited to model the beam degradation, collimation and passage through material.\ To overcome these limitations, TRANSPORT could be combined with TURTLE allowing the multi-particle tracking and with MUSCAT adding a basic particle-matter interaction routine to the tracking \cite{Turtle}.\ In the PROSCAN facility, the use of TRANSPORT together with TURTLE was limited to the development of the beam dynamics model toward Gantry 1 \cite{Rohrer2002}.\ The chain of codes TRANSPORT-TURTLE-MUSCAT resulted to be quite intricate to use since the parameters for the particle-matter interaction (e.g.\ scattering angle, average energy loss, absorption probability etc.)\ are computed by the auxiliary program MUSCAT and have to be manually included into the TURTLE input file.\ For the beamline toward Gantry 3 it has been decided to develop the model directly with OPAL.\ This code is equipped with a more accurate particle-matter interaction model that could be easier combined with the particle tracking.\ We expect that this more precise OPAL model, that includes the degrader, collimators and other scattering sources, will provide a better understanding of the beam properties and transmission along the beamline. 


\section{Multi-particle beam dynamics model in OPAL}
\label{sec:3}

OPAL (Object Oriented Particle Accelerator Library) is a three-dimensional tracker for general particle accelerator simulations \cite{OPAL}.\ It is based on the time integration of the equation of motion

\begin{equation}
	\frac{d\bm{p}}{dt} = q[\bm{E} + \bm{v}\times\bm{B}],
	\label{eq:Lorentz}
\end{equation}

where $\bm{p} = \bm{v}\cdot \gamma m_0$ is the momentum, $\bm{v}$ the velocity, $q$ the charge, $m_0$ the rest mass of the particle and $\gamma$ the relativistic factor.\ $\bm{E}$ and $\bm{B}$ indicate the electric and magnetic fields, respectively.\ OPAL uses the canonical variables $(x, p_x)$, $(y, p_y)$, $(z, p_z)$ and the time $t$ as independent variable to describe the trajectories of particles.\ Performing the calculations parallel on multiple processors, the tracking of each particle in the beam through the accelerator or beamline is obtained integrating \Eqref{eq:Lorentz} in a specific time step $\Delta t$, that corresponds to an equivalent $\Delta s$ in space.

The OPAL development began in 2008 with the initial purpose to simulate the particle orbits in cyclotrons with and without space charge.\ Lately, it has been extended to the particle tracking in other types of accelerators (e.g.\ linac and Fixed-Field-Alternating-Gradient synchrotrons) and beamlines.\ In 2013 the Monte Carlo model for particle-matter interaction was implemented in OPAL and it will be discussed in section \ref{subs:MC} \cite{Stachel2013}.\ In the following 4 years, the particle-matter interaction model has been improved, benchmarked, extended with new materials and connected with a GPU (Graphics Processing Unit) card that provides a remarkable speed-up to the computation time \cite{Locans2016}.\ The unique feature to combine seamless linear and nonlinear beam tracking  with Monte Carlo simulations of the particle-matter interaction makes OPAL a convenient code to model a proton therapy beamline.

The beamline elements are described in OPAL with some specific properties following the MAD convention \cite{Madx}.\ For the dipoles, for example, a field map including fringe fields is provided as default. The Enge function is used to model the entrance and exit fringe fields of the default field map \cite{Enge}.\ In addition, external field maps can also be loaded.\ In this way, the measured dipole fields can be simulated in OPAL and used in the tracking.


\subsection{Model setup with ROGER}
\label{subs:ROGER}

\begin{figure}[h!]
  \includegraphics[scale=0.17, keepaspectratio=true]{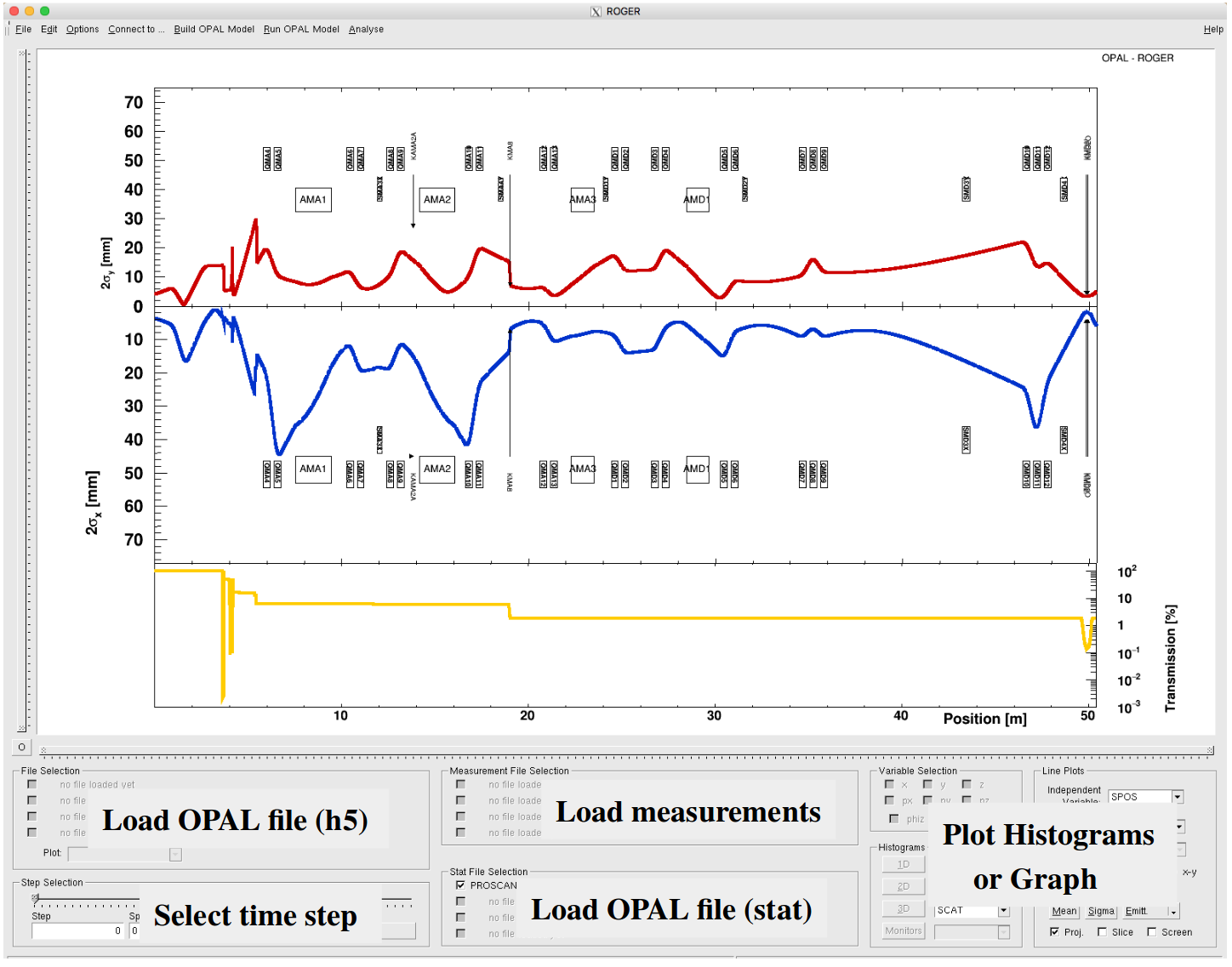}
    \caption{ROGER GUI and example of a data analysis.\ In the white canvas the beamline lattice, envelope (blue and red line for the horizontal-bending and vertical plane, respectively) and transmission (yellow line) are displayed.\ In the bottom part of the GUI, some basic ROGER functionalities (e.g.\ load files, plot options) are highlighted.}
    \label{fig:ROGER_GUI}
\end{figure}

To simplify the model setup, a direct connection between the PROSCAN database (containing the list and properties of the magnets, monitors, passive elements, etc.)\ and OPAL was convenient.\ For this purpose, ROGER (ROot GEnerator for Runopal) has been developed extending the H5root framework \cite{H5root}.\ Equipped with a dedicated GUI (\figref{fig:ROGER_GUI}) ROGER allows the users to easily build and run the OPAL beam dynamics model and to perform post-processing data analysis.

Once the connection with the database has been established, the lattice and the actual magnets settings of the selected transport line can be imported in ROGER and used to build the OPAL model.\ It is also possible to export the beamline settings in the format readable by the EPICS control system of the facility.

Due to the particle losses at the collimators ($\approx$ 90\%), an initial sample with a high number of protons ($10^6$ $-$ $10^7$) is needed in the simulation to collect enough statistics for the data analysis.\ As explained in section \ref{sec:3}, each particle of the sample is tracked by OPAL in a specific time step, by default set to 1 ps.\ The corresponding step in space ($\Delta s$) depends, of course, on the beam energy.\ For 1 ps time step $\Delta s$ varies from 0.1 mm to 0.04 mm in the energy range 230$-$70 MeV.  

After each time step, the entire particle sample and its mean values (e.g.\ beam size, momentum, emittance, etc.) are stored in the output files and are available for the data analysis. The use of a unique time step (e.g.\ 1 ps) for the entire beamline guarantees a higher model accuracy since the beam properties are stored within a fraction of a mm.\ At the same time, this leads to a longer simulation time ($\approx$ 1 hour for the full beamline toward Gantry 3) that is only partially reduced running in parallel on several processors.\ In addition such precision is not even required in some locations along the beamline.\ In the drift spaces and magnets, for example, a larger time step can be chosen without compromising the model accuracy.\ A shorter time step ($\approx$ ps) is needed in case of thin elements (e.g.\ monitors and collimators) as well as in case of particle-matter interaction simulation. In these cases, the use of a shorter time step is recommended since higher precision is required.\ However, as already mentioned, the Monte Carlo computation of the particle-matter interaction is performed by means of a dedicated GPU card.\ This allows boosting the performance of the Monte Carlo simulation even with a high number of initial protons and a shorter time step \cite{Locans2016}. 

In ROGER, the value of the time step can be set depending on the location and type of the element along the beamline.\ This adaptive time step solution allows reducing the simulation time to $\approx$ 10 minutes for the entire beamline without losing accuracy or precision.
 
The post-processing analysis is also performed in ROGER and the beam envelope, transmission (\figref{fig:ROGER_GUI}), beam profiles, losses at collimators resulting from the model are displayed and compared directly against the measurements.  


\subsection{Particle interaction with matter in OPAL}
\label{subs:MC}

One of the unique features of OPAL is to combine the particle tracking through an accelerator or beamline with a Monte Carlo simulation of the beam interaction with matter. 

In each time step in which a particle hits a material, OPAL performs a Monte Carlo simulation that calculates energy loss and elastic scattering.\ The amount of material $\Delta s$ that each particle of the beam crosses is related to the initial momentum of the particle and to the time step $\Delta t$ set in the OPAL input file.

It has to be remarked that the particle-matter interaction in OPAL is restricted only to protons and that the inelastic nuclear interactions are not implemented.\ The OPAL implementation of the proton interaction with matter is described in the next sections.\

\subsubsection{Energy loss}
\label{subsub:Bethe}

The amount of energy lost by a proton traveling through a material of thickness $\Delta s$ in a time step $\Delta t$ is given by
\begin{equation}
  dE_{\text{loss}} = d\overline{E} + \xi,
  \label{eq:ElossTot}
\end{equation}
where $d\overline{E}$ is the average energy loss and $\xi$ is a random number extracted from a random Gaussian distribution of width \cite{Leo}
\begin{equation}
  \sigma^2_E = Km_ec^2\rho\frac{Z}{A}\Delta s,
  \label{eq:sigma}
\end{equation}
where $K=4\pi N_Ar_e^2m_ec^2$, $N_A$ is the Avogadro's number, $r_e$ the classical electron radius, $m_e$ the electron mass and c the speed of light.\ The atomic number of the material, its atomic mass and density are specified by $Z$, $A$ and $\rho$ respectively.\

\Eqref{eq:sigma} represents the width of the Gaussian distribution that approximates the average energy loss in case of a relatively thick material where the number of interactions is large.\ With this approximation, the OPAL model neglects the asymmetric tail of this distribution which occurs at very large energy loss. The average energy loss $d\overline{E}$ in \Eqref{eq:ElossTot} is given by
\begin{equation}
 d\overline{E} = -\left\langle \frac{dE}{dx} \right\rangle \rho \Delta s,
 \label{eq:averE}
\end{equation}

where $-\left\langle dE / dx \right\rangle$ is the electronic stopping power.\ Following the ICRU (International Commission on Radiation Units and Measurements) 49 guideline \cite{ICRU}, the stopping power is defined by two different models according to the initial momentum of the incident particle.\ At energies lower than 0.6 MeV, the electronic stopping powers are obtained from experimental data and from the empirical formula developed by Andersen and Ziegler \cite{Andersen}.\ For energies higher that 0.6 MeV, the Bethe-Bloch equation is used \cite{PDG}.\ For a matter of completeness and generalisation, both models for high- and low- proton energies are implemented in OPAL. However, for the proton therapy application, the low-energies model can be neglected.  

The Bethe-Bloch equation implemented in OPAL is
\begin{equation}
  -\left\langle \frac{dE}{dx} \right\rangle = Kz^2 \frac{Z}{A}\frac{1}{\beta^2}\left[\frac{1}{2}\ln\left(\frac{2m_ec^2\beta^2\gamma^2W_{\text{max}}}{I^2}\right)-\beta^2\right],
  \label{eq:Bethe}
\end{equation}
where $z$ is the proton charge and $\beta$ and $\gamma$ are the relativistic kinematic factors.\ The mean excitation energy ($I$) depends on the material properties and OPAL uses the definitions given in \cite{Leo}.\ In \Eqref{eq:Bethe}, $W_{max}$ is the maximum kinetic energy that a free electron acquires in a single collision and is expressed by
\begin{equation}
  W_{\text{max}} = \frac{2m_ec^2\beta^2\gamma^2}{1+2\gamma m_e/m_p +(m_e/m_p)^2},
\end{equation}
where $m_p$ is the proton mass. 

The Bethe-Bloch (\Eqref{eq:Bethe}) does not contain the density-effect correction $\delta$.\ This factor describes the reduction of the stopping power due to the polarization of the medium and it is important only in the ultra-relativistic regime.\ Therefore, for proton therapy application, neglecting this term is perfectly reasonable. 

In each time step the same model for the energy loss is continuously applied to all protons in the beam.\ In addition, a proton is removed from the beam when its kinetic energy is less than 0.1 MeV. 

\subsubsection{Coulomb scattering}

\begin{figure}[h]
\centering
  \includegraphics[scale=0.10, keepaspectratio=true]{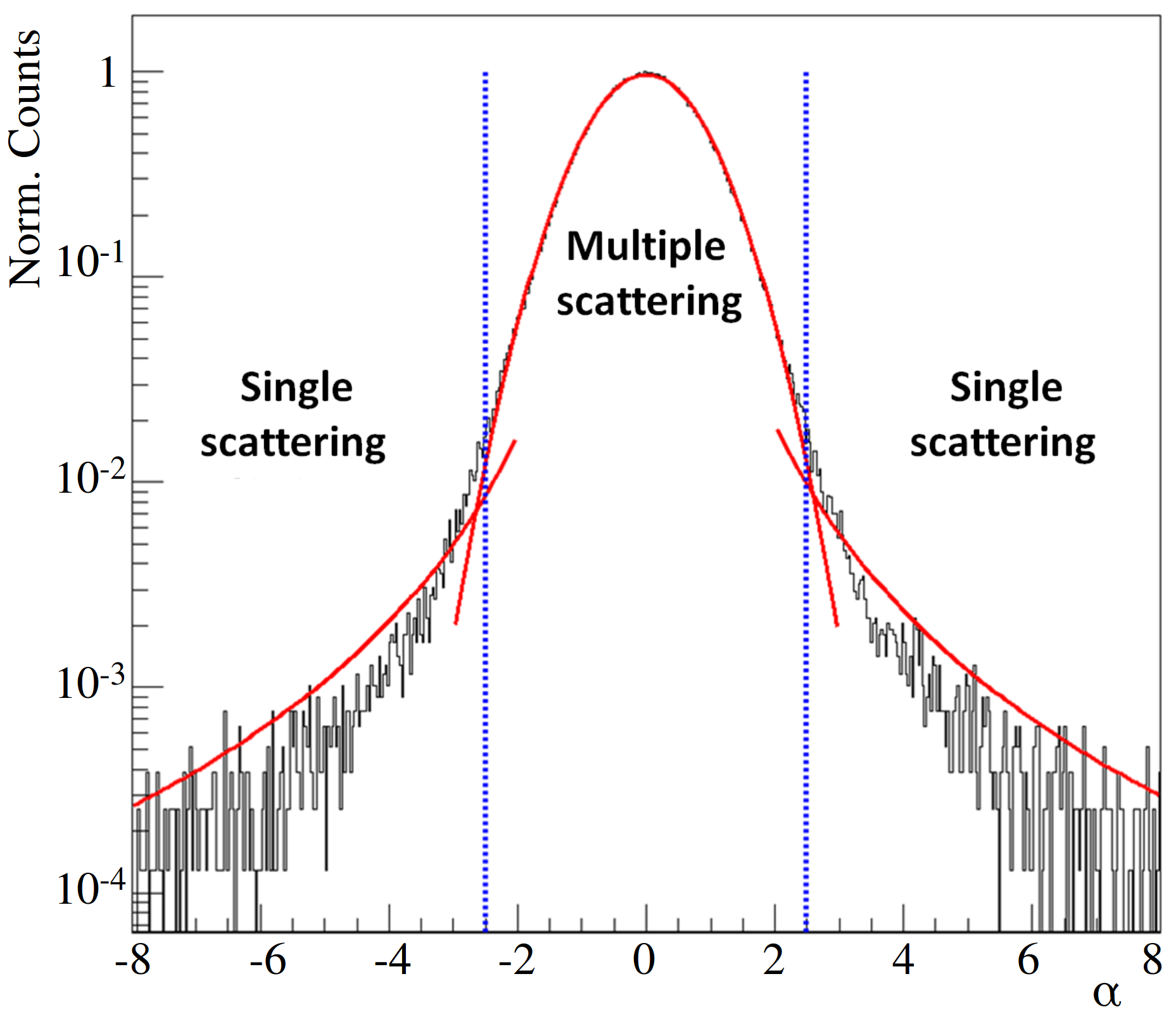}
    \caption{Angular distribution of a scattered proton beam from graphite in function of the projected angle $\alpha$ simulated in OPAL.\ The red curves are the fit from the Multiple and Single scattering distributions (\Eqref{eq:PM} and \Eqref{eq:PS}).\ The dashed-blue lines set the transition between Multiple and Single Scattering.} 
    \label{fig:Scat_distr}
\end{figure}

In OPAL a simplified Coulomb Scattering (CS)  model is available, following \cite{Jackson}.\
The trajectory of a proton that travels through a material of thickness $\Delta s$ (corresponding to a time step $\Delta t$) is altered due to many small deflections, referred to as Multiple Coulomb Scattering (MCS), or to a single large deflection, referred to as Single Rutherford Scattering (SRS).\ The relative projected angle $\alpha$ is used as to define a threshold between MCS and SRS. The transition from MCS to SRS occurs for
\begin{equation}
 \alpha = \frac{\theta_{\text{scat}}}{\sqrt{\langle\Theta ^2\rangle}} = 2.5,
 \label{eq:alpha}
\end{equation}
where $\theta_{\text{scat}}$ is the angle of scattering and $\langle\Theta ^2\rangle$ is the mean square angle.\ The Multiple and Single scattering distributions in terms of the projected angle are
\begin{equation}
	P_{\text{Multiple}}(\alpha) d\alpha = \frac{1}{\sqrt{\pi}}e^{-\alpha^2} d\alpha \mbox{ \hspace{0.2cm} and}
	\label{eq:PM}
\end{equation}
\begin{equation}
 	P_{\text{Single}}(\alpha) d\alpha = \frac{1}{8\cdot \ln{(204Z^{-1/3})}}\frac{d\alpha}{\alpha^3},
 	\label{eq:PS}
\end{equation}
where Z is the atomic number of the material \cite{Jackson}.\ Combining \Eqref{eq:PM} and \Eqref{eq:PS}, the complete angular distribution shows a Gaussian core due to the MCS with lateral tails due to the SRS (\figref{fig:Scat_distr}).

In terms of $\theta_{\text{scat}}$, the transition from MCS to SRS occurs for $\theta_{\text{scat}} = 3.5$ $\theta_0$, where $\theta_0$ is the scattering angle from Moliere's theory given by
\begin{equation}
 \theta_0 = \frac{13.6 \text{~MeV}}{\beta c p}z\sqrt{\frac{\Delta s}{X_0}}\left[1+0.038\ln{\left(\frac{\Delta s}{X_0}\right)}\right],
 \label{eq:Moliere}
\end{equation}
where $X_0$ is the radiation length of the material.

For each proton in the sample crossing a material OPAL evaluates \Eqref{eq:Moliere}.\ Then $\theta_0$ is multiplied by a random number from a Gaussian distribution of mean 0 and width 1 obtaining $\theta_{\text{scat}}$.

If $\theta_{\text{scat}}$ is lower than 3.5 $\theta_0$, then the particle undergoes MCS with an angle $\theta_{\text{MCS}}$ equal to $\theta_{\text{scat}}$.\ If $\theta_{\text{scat}}$ is larger than 3.5 $\theta_0$, then the particle belongs to the SRS tails of the angular distribution. The corresponding $\theta_{\text{SRS}}$ angle is calculated from \Eqref{eq:Moliere} and from a second random number $\xi_2$ between 0 and 1 such that:
\begin{equation}
	\theta_{SRS} = \pm 2.5\cdot \sqrt{\frac{1}{\xi_2}}\cdot \theta_0,
\end{equation}
where the positive or negative sign is given by a third random number that determines the scattering direction (up- or downwards). 

Once the scattering angle $\theta_{\text{MCS}}$ or $\theta_{\text{SRS}}$  has been evaluated, the position and direction of the particle are consequently updated following the notation of \cite{PDG}. 


\section{Results}
\label{sec:4}

In the next sections the OPAL model of the beamline toward Gantry 3 is discussed.\ For several energies in the range 230$-$70 MeV, the OPAL model was prepared and validated against different types of measurements performed during the commissioning of the beamline. 


\subsection{Degrader simulation and energy calculation}
\label{subsec:degrader}

\begin{figure}[h!]
	\subfloat[Real layout: wedges \cite{VanGoethem2009}]{\includegraphics[scale=0.22, keepaspectratio=true]{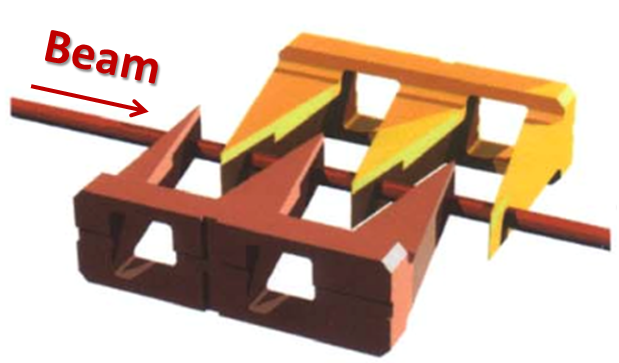}\label{fig:degrader}} 
	\subfloat[OPAL geometry: slabs]{\includegraphics[scale=0.12, keepaspectratio=true]{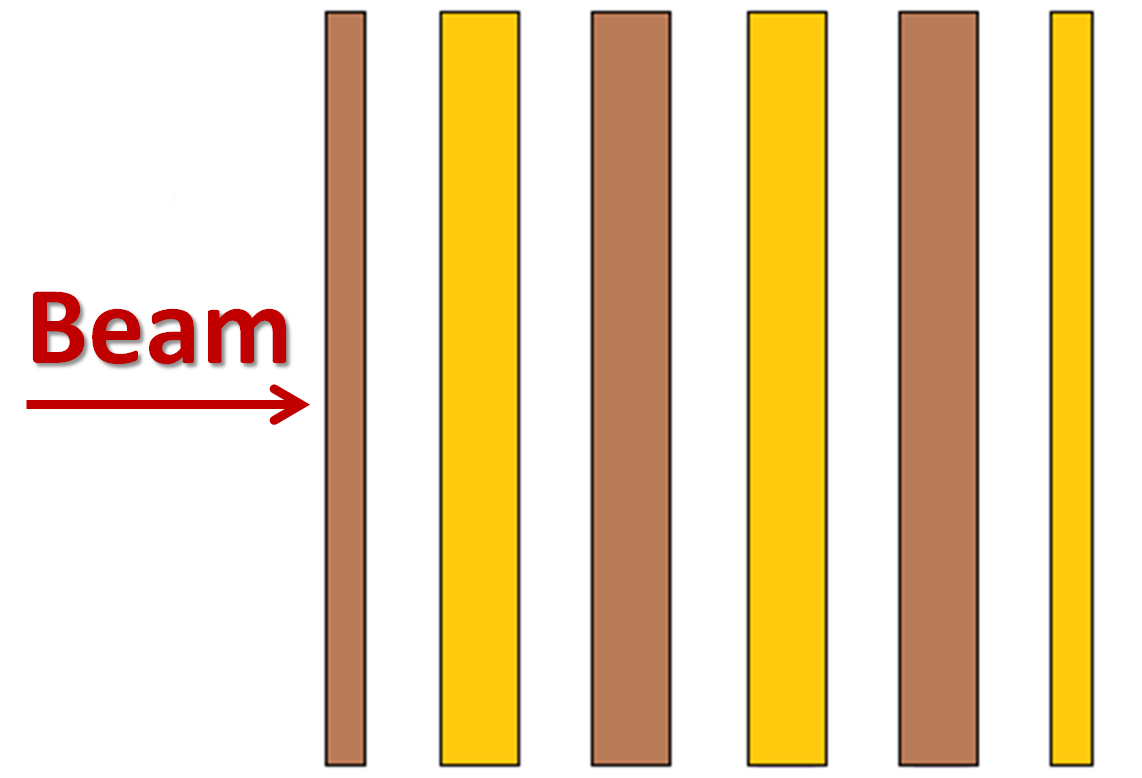}\label{fig:Deg_slab}}
	\caption{PROSCAN degrader.}
\end{figure}

In a cyclotron-based facility, the changes in energy needed to scan the tumor in depth are performed by means of a degrader.\ At PROSCAN, this device consists of 2 pairs of 3 movable wedges of graphite, as shown in \figref{fig:degrader}.\ In less than 50 ms, the wedges move increasing or reducing the thickness of graphite that the beam encounters.\ The selected energy in the range of 230$-$70 MeV is delivered with an accuracy of $\pm 0.1$ mm water-equivalent \cite{Reist2002}.\ Since in OPAL the wedge geometry of the degrader can not be recreated, a simplified geometry is implemented (\figref{fig:Deg_slab}) and its impact on the 250 MeV incoming beam from Comet has been modelled using the particle-matter interaction models described in section \ref{subs:MC}.

The wedge geometry of the degrader was simulated with FLUKA \cite{Fluka} using the configuration at 230 MeV, where the wedge shape is expected to have a bigger influence.\ From the comparison between slab and wedge geometry no significant differences in the beam properties were found.\ This result validates the simplification adopted in the OPAL geometry.

\begin{figure}[h!]
\centering 
  \includegraphics[scale=0.12, keepaspectratio=true]{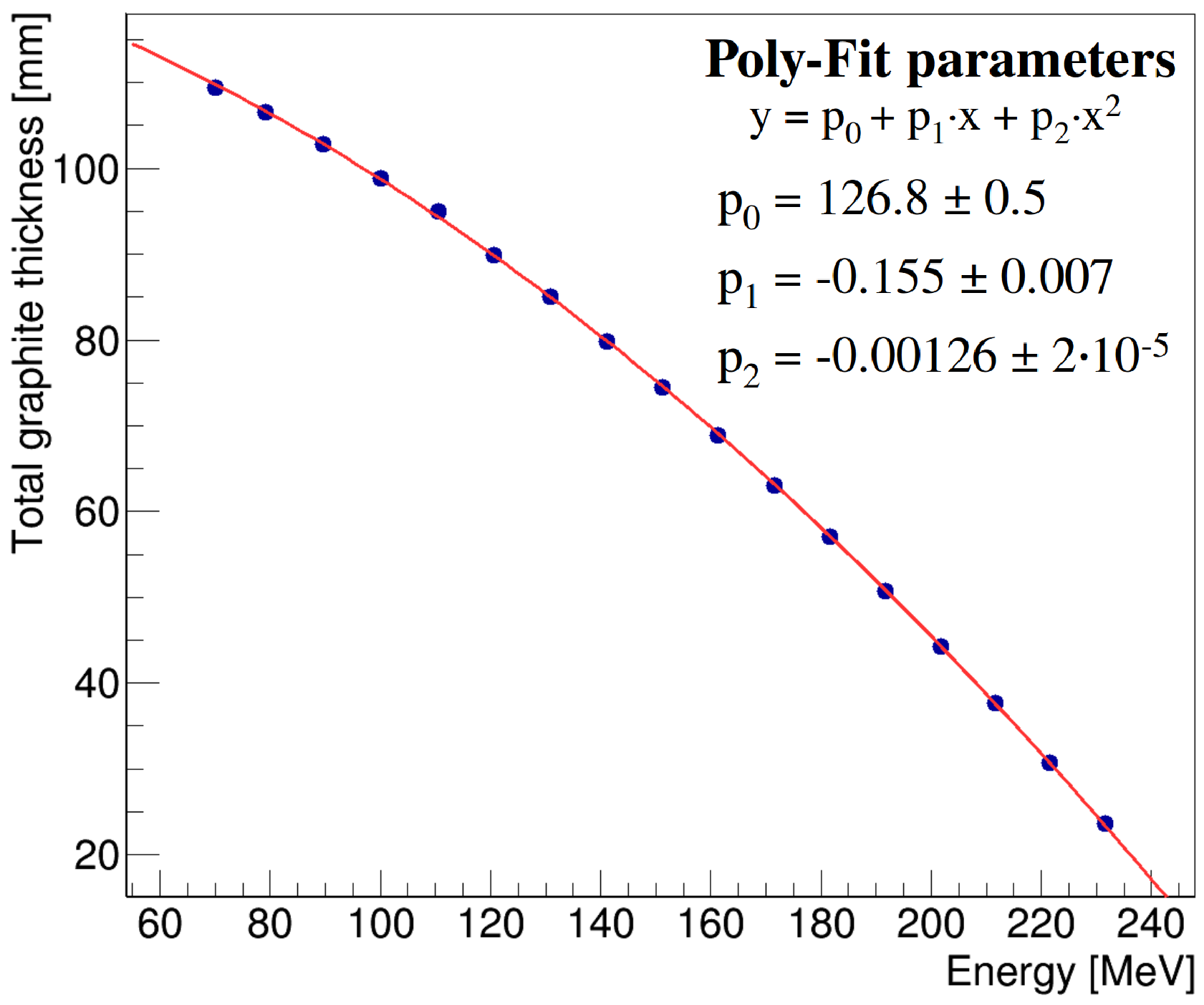}
    \caption{Calibration curve for the degrader: the total graphite thickness as a function of required beam energy.}
    \label{fig:deg_calib}
\end{figure}

The thickness of the 6 slabs is adjusted depending on the required final energy, according to the calibration curve of \figref{fig:deg_calib}.\ 

As mentioned in section \ref{subs:ROGER}, the Monte Carlo simulation of the proton beam interaction with the degrader is performed by means of a GPU card.\ The obtained speedup is x140$-$x160 compared to the time required for the same computation without GPU \cite{Locans2016}.\ This allows keeping a high precision as required for the degrader simulation without increasing the overall computation time. 

\begin{figure}[h]
	\centering
	\subfloat[Degrader setting: 230 MeV]{\includegraphics[scale=0.125, keepaspectratio=true]{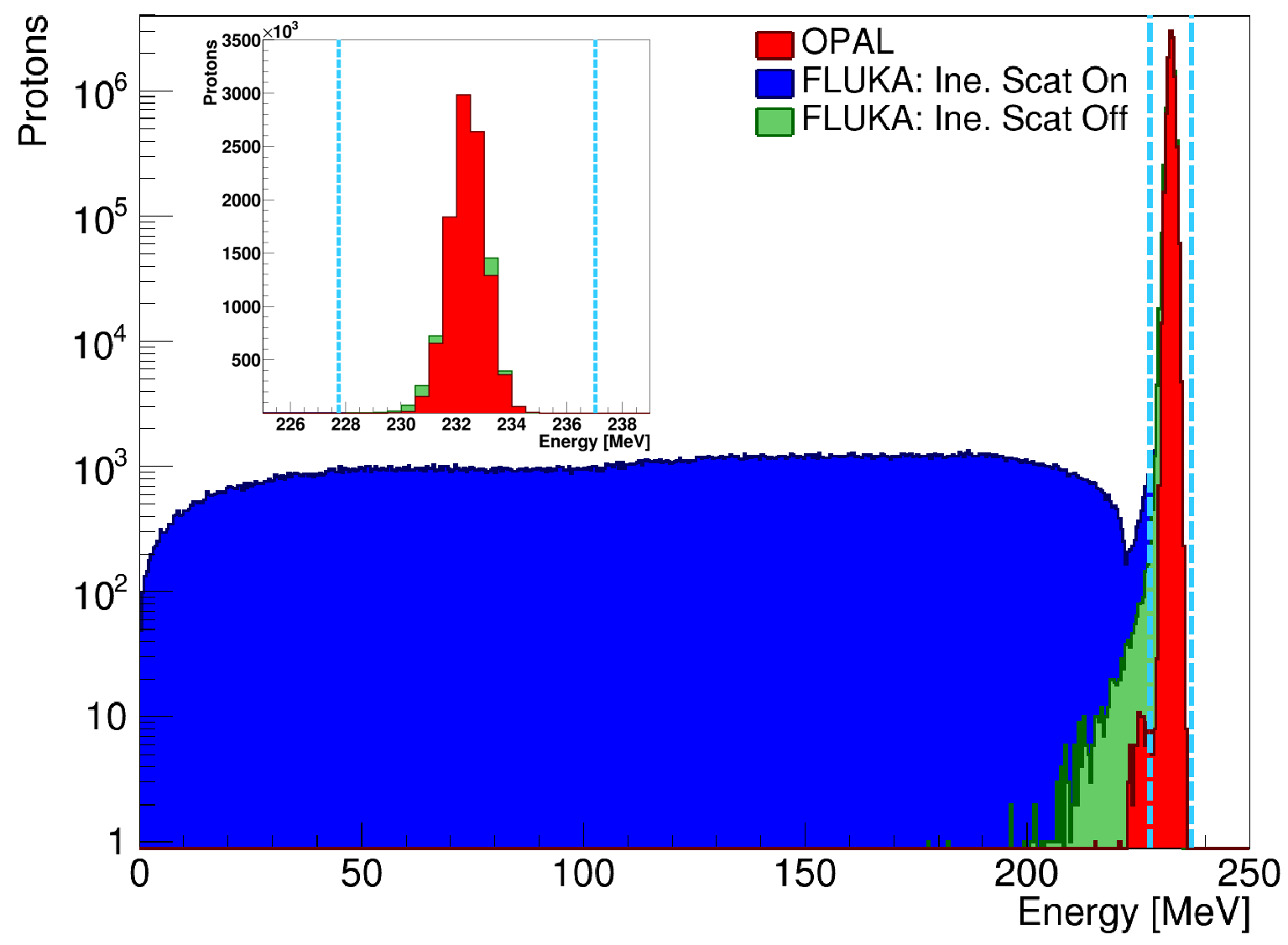}\label{fig:DMAD_230}}\\
	\subfloat[Degrader setting: 70 MeV]{\includegraphics[scale=0.12, keepaspectratio=true]{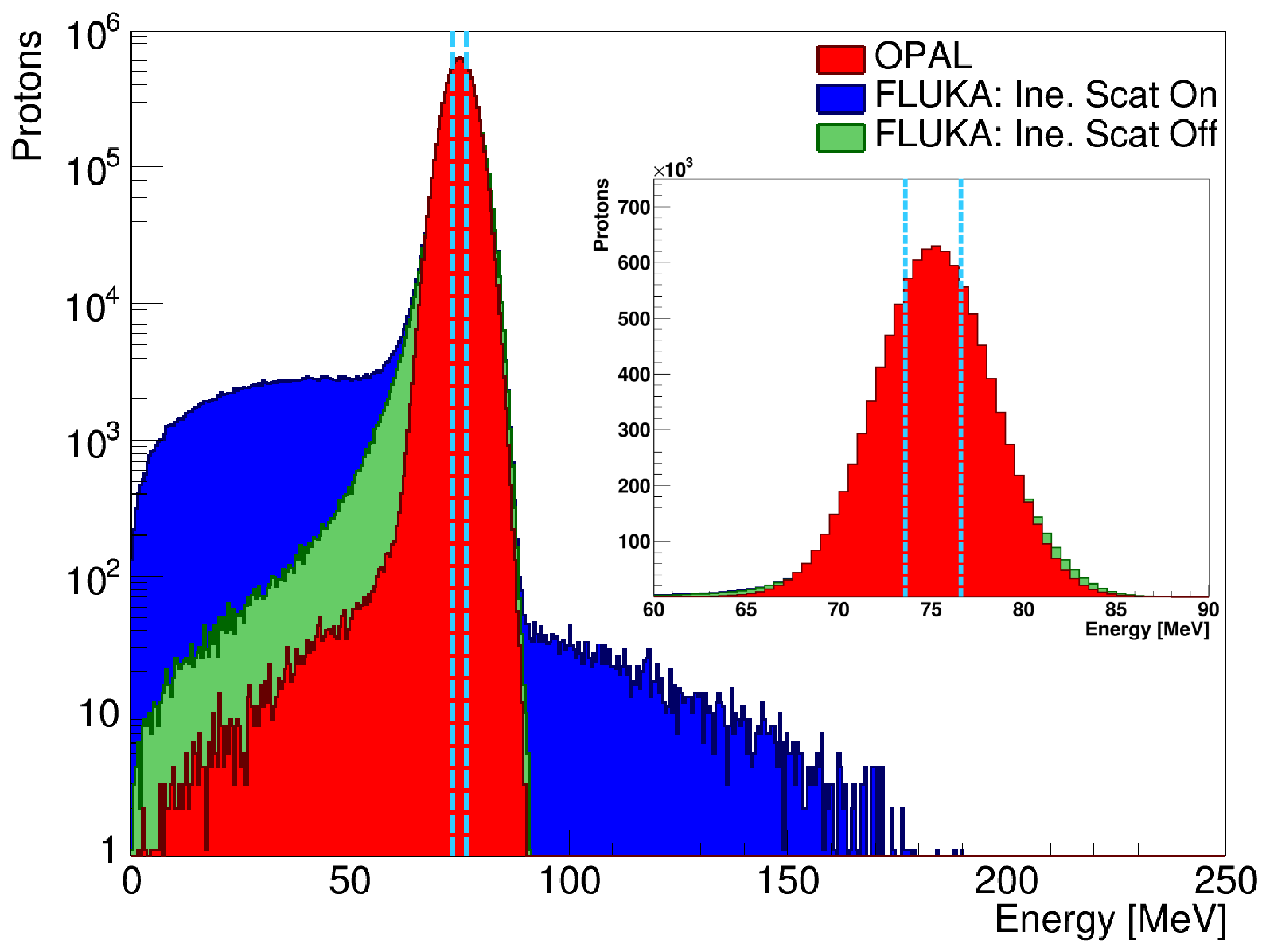}\label{fig:DMAD_70}}
	\caption{Energy distribution of the proton beam after the interaction with the degrader.\ The OPAL results are compared with FLUKA with and without the inelastic scattering contribution.\ The light-blue
 dashed lines reproduce the energy acceptance width of the ESS.\ In the zoom the overlap between the energy peak and the acceptance of the ESS is shown in more detail.}
	\label{fig:OPAL_Fluka_Scattering}
\end{figure}

The OPAL model for different degrader configurations (or energies) was benchmarked against FLUKA.\ \figref{fig:OPAL_Fluka_Scattering} shows two examples of the proton energy distribution resulting from OPAL and FLUKA.\ In particular, the thickness of 6 graphite slabs was set to reduce the beam energy from 250 MeV to 230 MeV (\figref{fig:DMAD_230}) and to 70 MeV (\figref{fig:DMAD_70}).\ The main discrepancy between the two codes is due to the inelastic scattering contribution that is not available in OPAL.\ Disabling the inelastic scattering also in FLUKA, a better agreement was found on the mean energy described by the Bethe-Bloch in \Eqref{eq:Bethe}.  

The reduced OPAL accuracy excluding the inelastic scattering is, for this application, almost irrelevant.\ In \figref{fig:OPAL_Fluka_Scattering} the two light-blue dashed lines indicate the fraction of the energy distribution that will be selected by the ESS, reducing to $\pm 1.0\%$ the momentum spread of the degraded beam.\ The inelastic tails of the distribution are hence removed by the horizontal slit in the ESS.  


\subsection{Energy calculation and measurement}
\label{subsec:energy}

In the transport line toward Gantry 3, the beam energy can be measured at three different locations.\ A first energy measurement can be performed with the monitor in the dispersive area of the ESS using the first dipole after the degrader (AMA1 in \figref{fig:Envelope}) as a spectrometer.\ Downstream the Intensity Compensation Collimator (ICC), it is possible to install a multi-leaf Faraday cup and measure the energy by the proton range \cite{Doelling2015}.\ Before the installation of Gantry 3, a range measurement with a water tank was performed at the end of the fixed beamline.

\begin{figure}[h!]
  \includegraphics[scale=0.16, keepaspectratio=true]{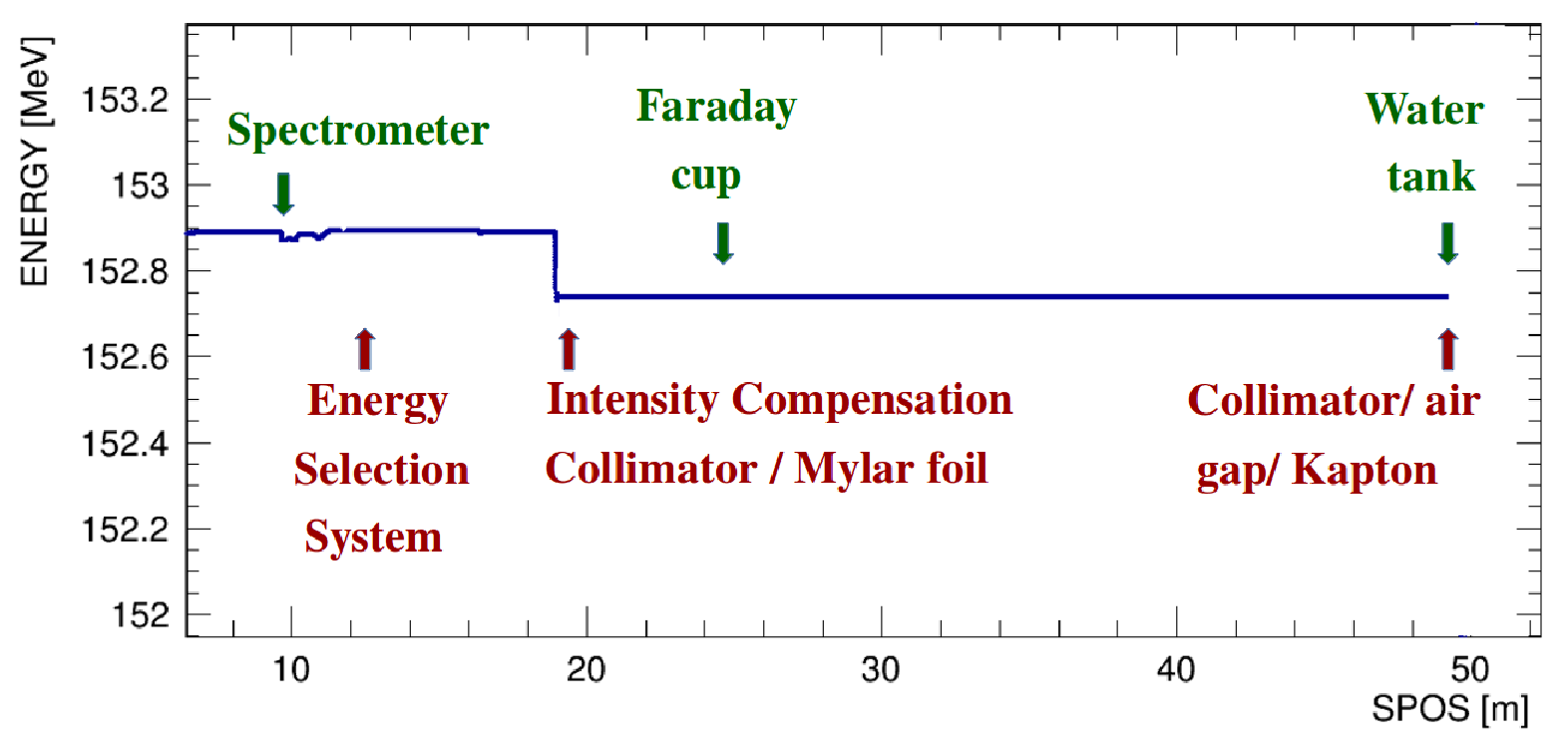}
    \caption{Average beam energy along the beamline from the degrader to the coupling point of Gantry 3.\ The causes of the energy reduction are highlighted in red, while in green the devices used for the energy measurements.}
    \label{fig:EnergyLoss_150}
\end{figure}

The momentum cut from the horizontal slit in the ESS and the interaction with the ICC, scattering foils and air gaps at the coupling point change slightly the average beam energy along the beamline.\ The energy loss is around 0.2$-$0.3 MeV. This has been verified in simulation with the OPAL model, as shown in \figref{fig:EnergyLoss_150}. 

Performed at the end of the fixed transport line, the measurements with the water tank reveal the beam energy entering the Gantry 3.\ In \tabref{table:Energy}, the results from the OPAL model are compared against the measured energies extrapolated from the ICRU conversion of proton range in water \cite{ICRU}.\ Averaged over the entire energy range, the discrepancy between the OPAL model and the measurements is less than 0.2\%, which is reasonable for this application.\ This agreement validates the particle-matter interaction model in OPAL and the calibration of the slab thickness in relation to the reduced energy in the degrader model (see \figref{fig:deg_calib}).  

\begin{table}[ht!]
\caption{Mean energy in MeV at the coupling point of Gantry 3 for different degrader settings: comparison between the Monte Carlo simulation from OPAL and the extrapolated energy from the proton range measurements with the water tank.}
\begin{ruledtabular}
 	\begin{tabular}{lcr}
		\textbf{Setting} & \textbf{OPAL} & \textbf{Measurements} \\	
		\hline
		\hline
		230 & 231.69 & 231.20 $\pm$ 0.17 \\
		210 & 211.82 & 211.29 $\pm$ 0.17 \\
		190 & 192.10 & 191.61 $\pm$ 0.16 \\
		170 & 172.22 & 171.81 $\pm$ 0.16 \\
		150 & 152.13 & 152.05 $\pm$ 0.16 \\
		110 & 112.46 & 112.17 $\pm$ 0.16 \\
		90 &  92.30 & 92.18 $\pm$ 0.16 \\
		70 & 73.66 & 73.88 $\pm$ 0.17 \\
	\end{tabular}
\end{ruledtabular}

\label{table:Energy}
\end{table}


\subsection{Envelope and transverse beam profiles}

\begin{figure*}[htb]
	\centering
		\subfloat[From degrader to the ICC]{
			\includegraphics[scale=0.107, keepaspectratio=true]{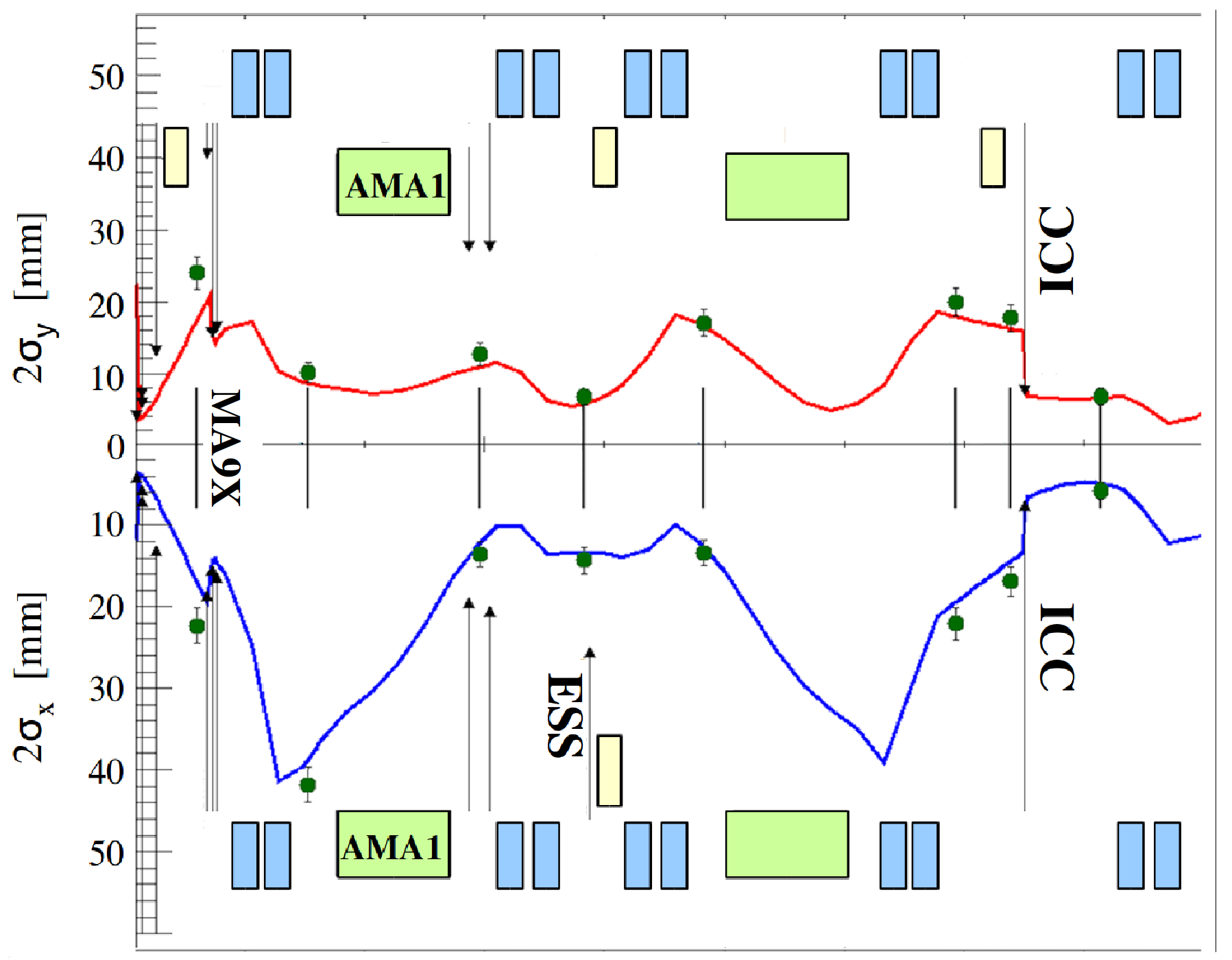}\label{fig:Envelope1}
			} 
		\subfloat[From the intensity compensation collimator to the coupling point]{
			\includegraphics[scale=0.13, keepaspectratio=true]{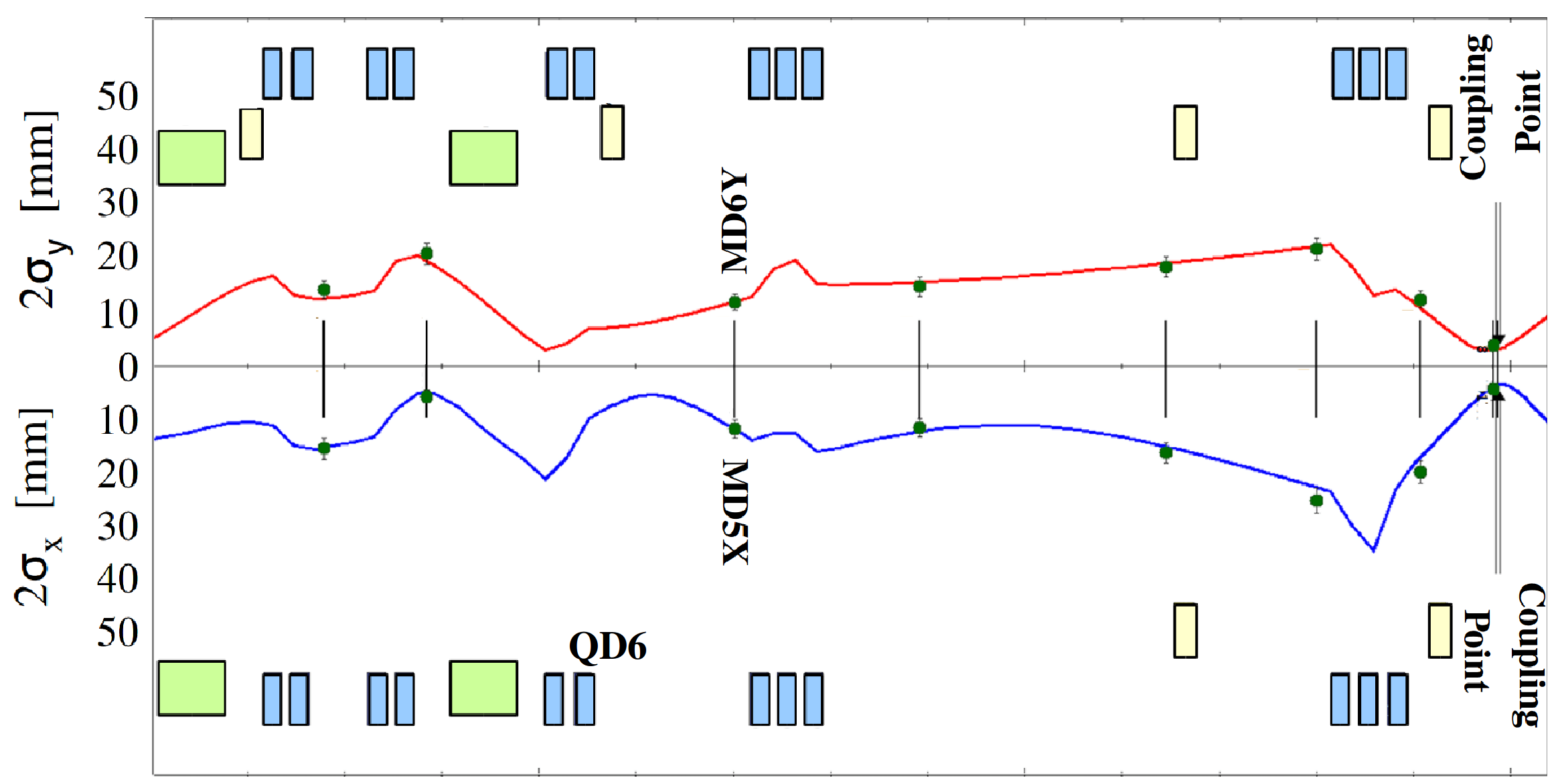}\label{fig:Envelope2}		
			}
	\caption{2$\sigma$ envelope at 230 MeV from the degrader up to the coupling point of the Gantry 3.\ The red and blue lines represent the beam size from OPAL in the vertical and horizontal-(bending) plane, respectively.\ The green dots represent the measured 2$\sigma$ beam size after a cut of the lateral tails of the measured beam profile below 10\%.\ The lattice is also drawn, in particular: monitors (vertical black lines), collimators (vertical arrows), dipoles (light green box), quadrupoles (light blue box) and steering magnets (light yellow box).}
	\label{fig:Envelope}
\end{figure*}

Using the post-processing analysis tool in ROGER, the 2$\sigma$ beam envelope resulting from the OPAL model can be displayed together with the beamline lattice, as shown in \figref{fig:Envelope}.\ A direct comparison with the measured beam sizes is also possible allowing an immediate validation of the model.

\begin{figure}[h!]
		\subfloat[Larger RMS discrepancy]{\includegraphics[scale=0.10, keepaspectratio=true]{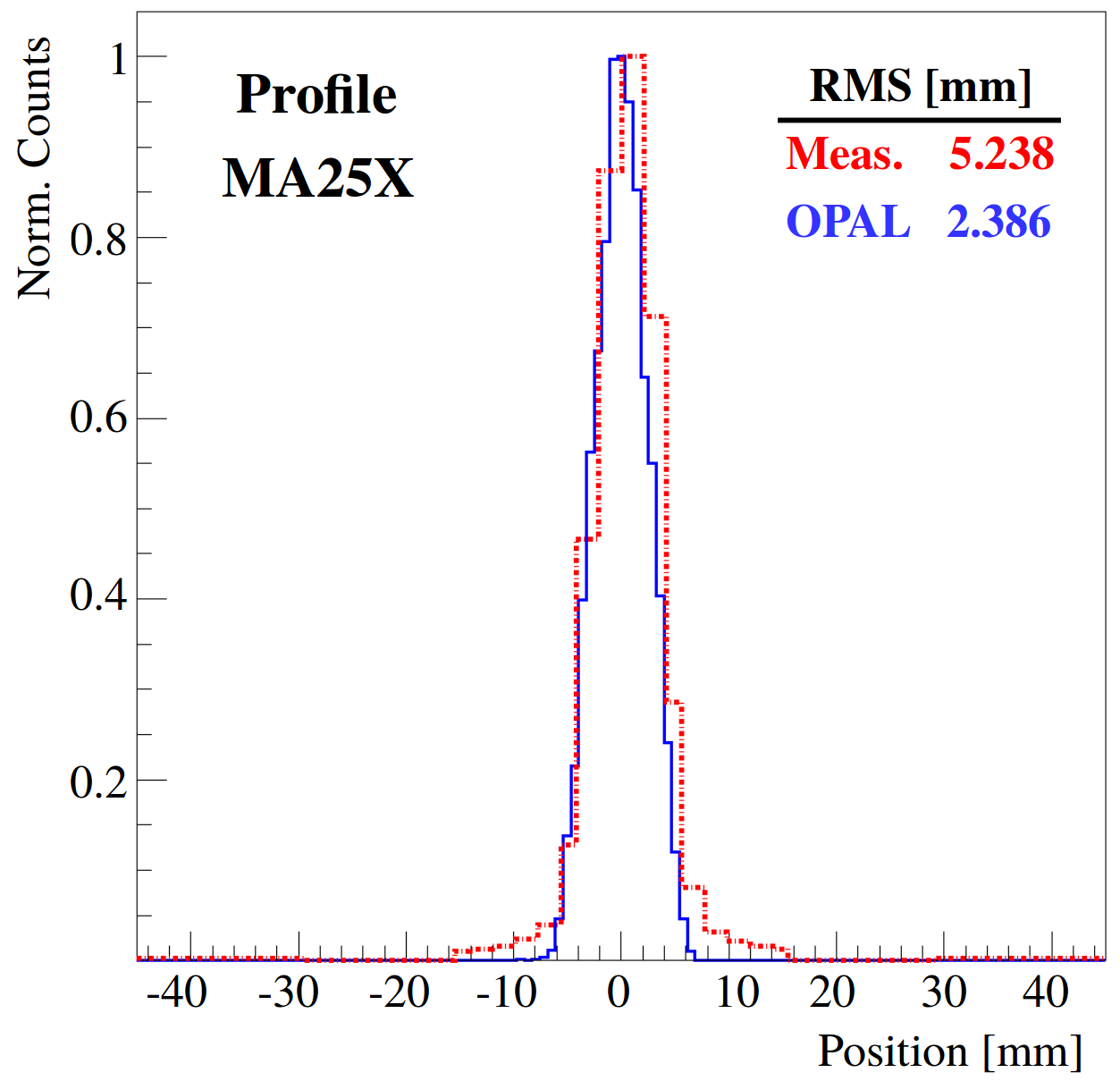}\label{fig:large}} 
		\subfloat[Smaller RMS discrepancy]{\includegraphics[scale=0.10, keepaspectratio=true]{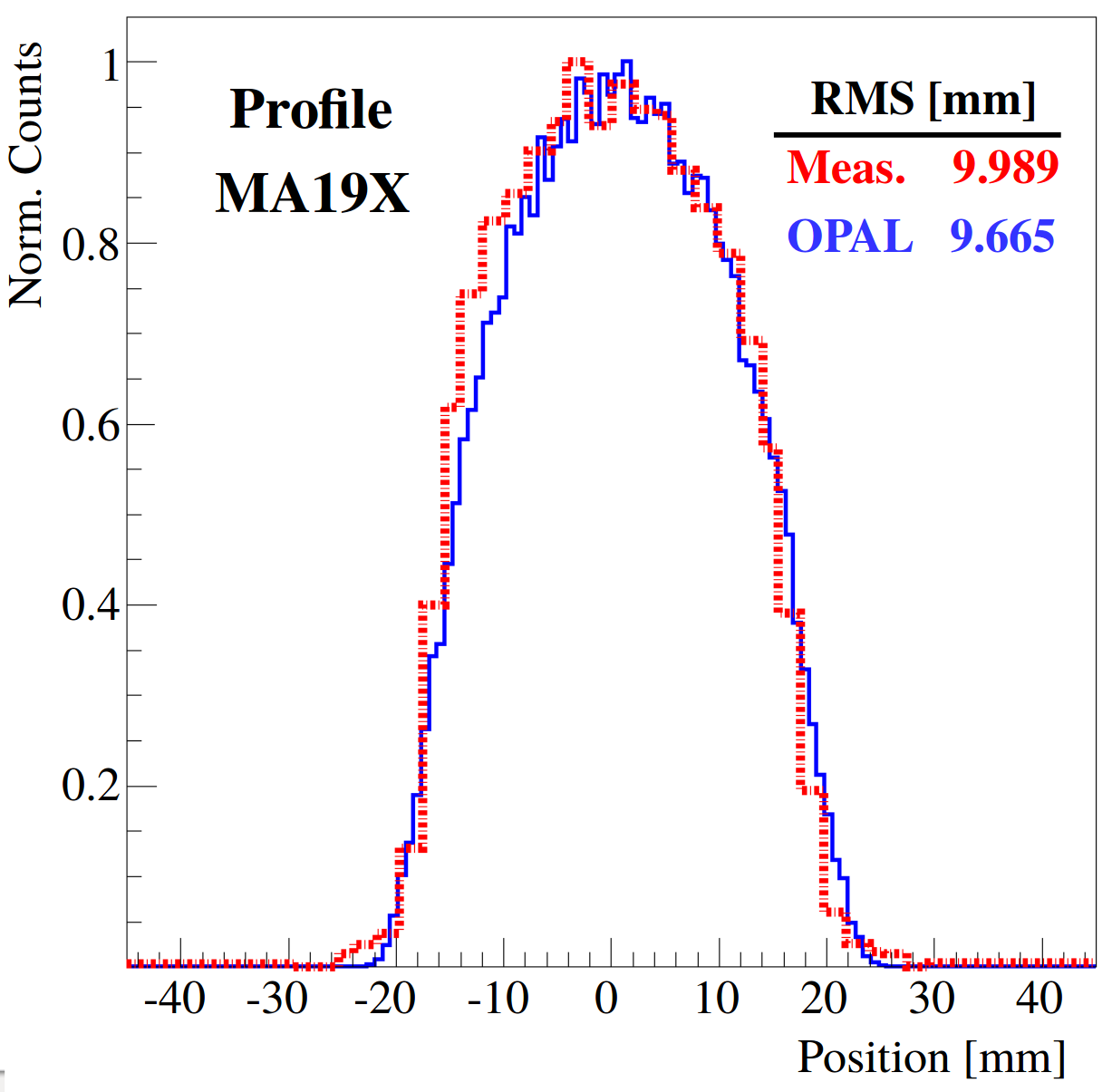}\label{fig:small}}
	\caption{Comparison between the particle distribution from the OPAL model with the measured beam profiles from two selected monitors: MA25X placed after a collimator and MA19X placed in a free area away from a collimator.}
	\label{fig:profile}
\end{figure}

The beam profile measurements along the beamline are performed with 21 pairs of retractable strip monitors.\ These devices measure beam position, transversal profile and, with a proper calibration, also the absolute beam current \cite{Doelling2012}.\ A preliminary analysis is performed on the measured beam sizes and the results are displayed in \figref{fig:Envelope} (green dots).\ In this preliminary analysis, a cut of the tails below 10\% is applied on all normalised measured profiles.\ This threshold includes the contribution of electronic noise and scattered particles on the measured beam size.\ Lately, a more precise analysis was performed comparing, for each monitor, the particle distribution from the OPAL model with the measured profile.\ This allows a better understanding of the contribution of noise or scattering to the measured beam size.\ An example is shown in \figref{fig:profile} for two particular monitors in the beamline: monitor MA25X (\figref{fig:large}) placed after a collimator and monitor MA19X (\figref{fig:small}) placed in a free area before a collimator.

Besides the qualitative agreement of \figref{fig:large} between the measured and the simulated profile, the quantitative difference in the RMS is almost 3 mm.\ The beam interaction with the collimator ahead of MA25X produces scattered particles that create tails of the measured profile. This leads to a discrepancy between the simulated and measured RMS beams size larger than 20\%.\ The same discrepancy is found at all monitors located sharply behind a collimator.\ For other monitors, such as \figref{fig:small} distant from the collimators, this discrepancy is less than 10\%, validating the OPAL results.


\subsection{Scattering effect from the collimators}
\label{subsec:ScatOnCol}

The results of \figref{fig:large} reveals the importance to include in the optics model the scattering effect from the collimators.\ At high energies, the scattered particles create tails on the measured profiles resulting in an enlarged discrepancy with the prediction from the simulation, if not properly modelled \cite{VanLuijk2001}.\ For this reason, the particle-matter interaction model of section \ref{subs:MC} was applied not only to the degrader, but also to the collimators of the beamline. 
 
\begin{figure}[h!]
  \subfloat[Without Monte Carlo]{\includegraphics[scale=0.10, keepaspectratio=true]{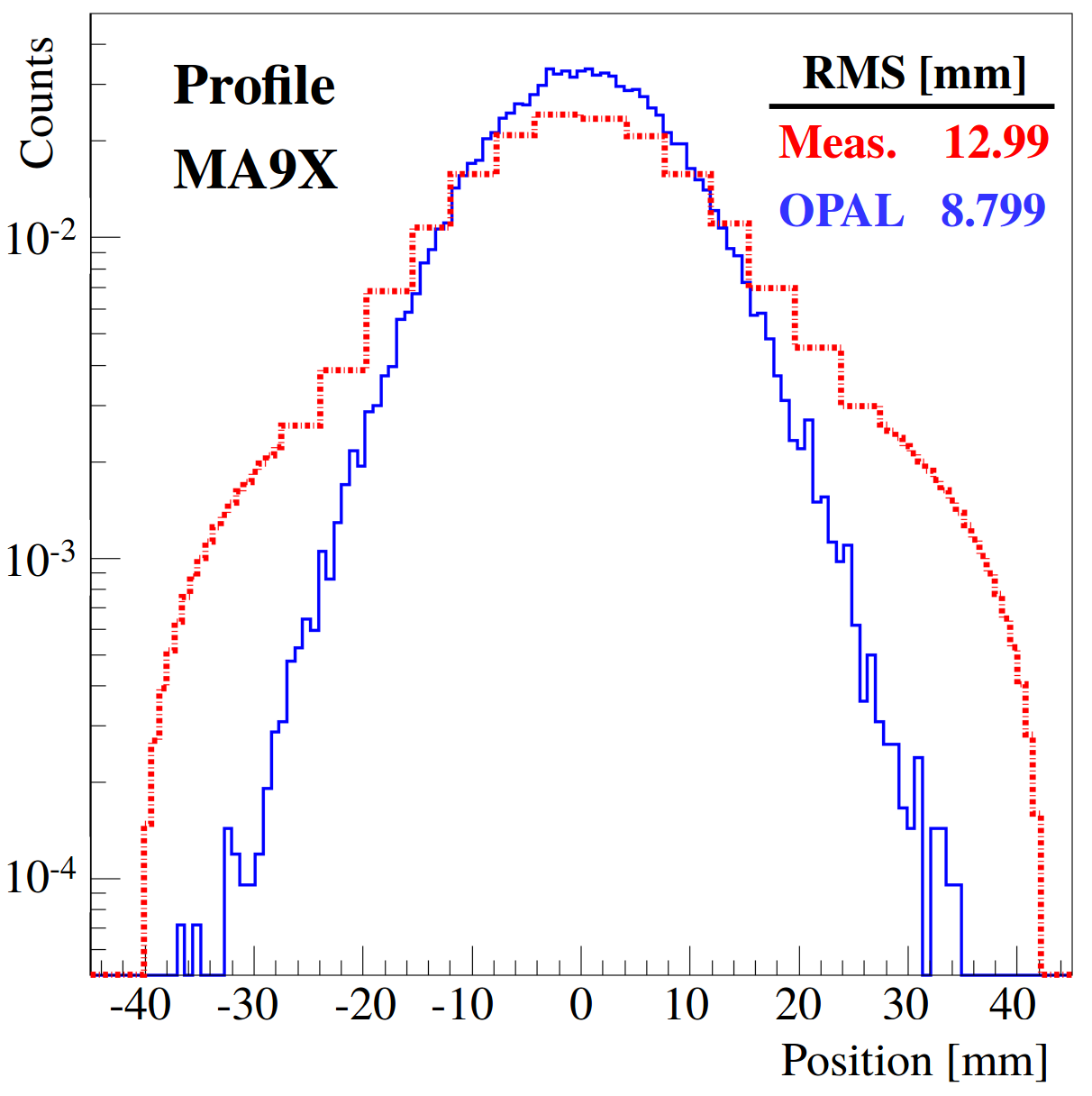}\label{fig:NoScat}} 
  \subfloat[With Monte Carlo]{\includegraphics[scale=0.10, keepaspectratio=true]{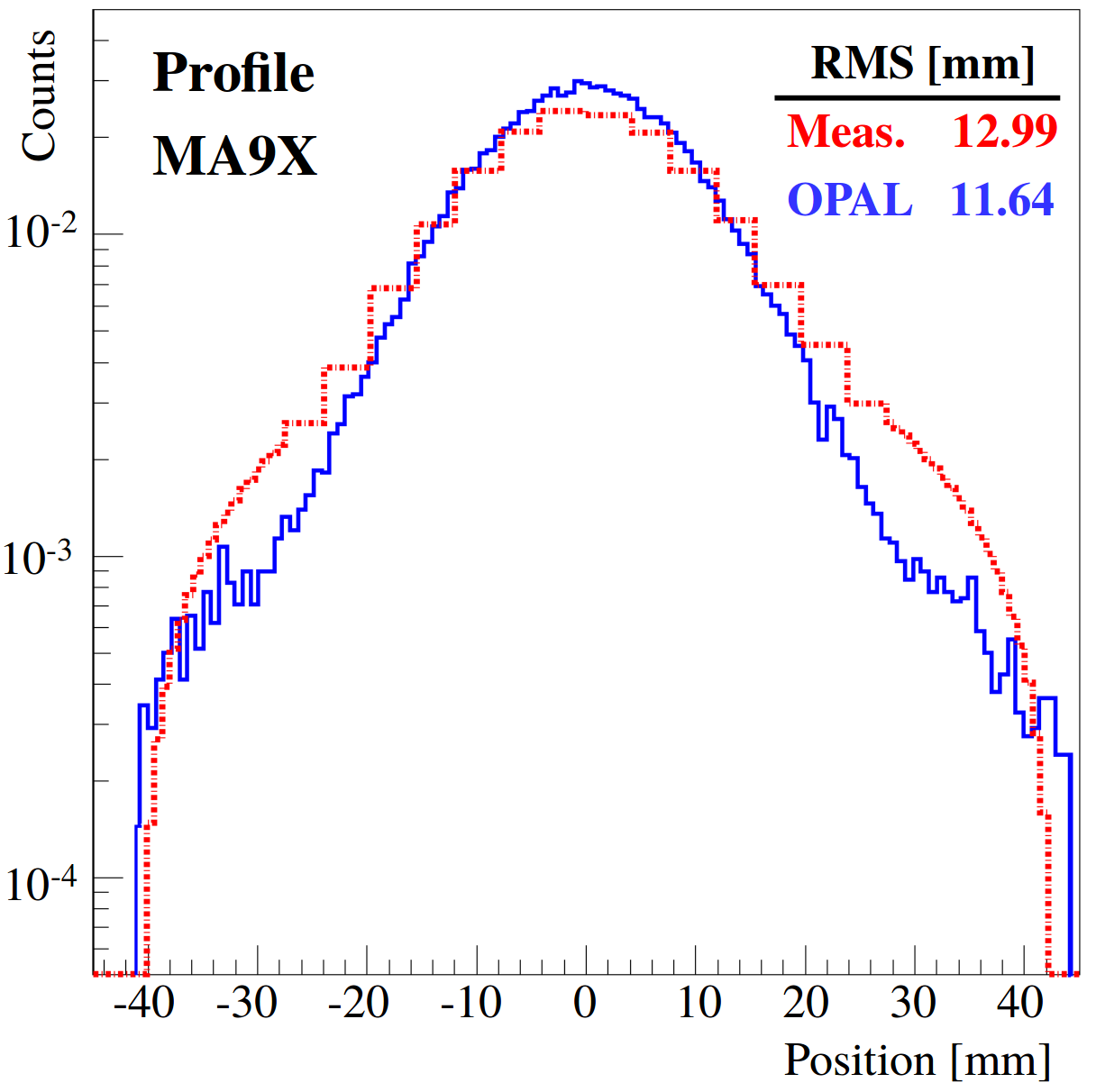}\label{fig:Scat}}
  \caption{Comparison between the measured profile and the particle distribution from OPAL with and without the Monte Carlo routine for the particle-matter interaction applied on the collimators ahead of the monitor MA9X.}
  \label{fig:CollimScat}
\end{figure}

Besides the collimator in front of monitor MA25X, the main collimation of the beam takes part right after the degrader by means of three consecutive collimators.\ The first collimator (KMA3) is made of copper, has variable circular aperture and is used to define the size of the beam along the transport line.\ The second collimator (KMA4), made of carbon, has a larger fixed aperture and absorbs the particles that bypassed KMA3.\ The third collimator (KMA5) is also made of copper, has variable apertures and is used to limit the beam divergence along the transport line.\ The profile monitor MA9X (see \figref{fig:Envelope1}) is placed after the first two collimators.\ At higher energies, the beam distribution recorded by this monitor shows two lateral tails due to the scattered particles from KMA3 (mainly) and KMA4 (see \figref{fig:CollimScat}).\ In the OPAL model, these two collimators have been defined with proper aperture, length and material (copper and carbon) and the particle tracking performed including the particle-matter interaction models.\  The resulting particle distribution for the highest energy (230 MeV) is shown in \figref{fig:CollimScat} in comparison with the measured profile.\ As visible in \figref{fig:NoScat}, the measured profile appears to be very broad.\ The external smooth shoulders are artifacts arising from the interpolation procedure of the strips signal.\ With the particle-matter interaction models applied on the collimators (\figref{fig:Scat}), an improvement of almost 30\% was accomplished on the modelled RMS beam size.

This analysis underlines the importance of combining Monte Carlo simulation with particle tracking.\ OPAL offers this possibility in a single code.\ The same analysis requires in general the use of at least two codes \cite{VanGoethem2009, Pavlovic2008}.


\subsection{Transverse emittance}
\label{sec:emit}

The multiple scattering of the protons by the degrader increases the beam emittance beyond the acceptance of the beamline.\ As explained before, the degraded beam is then collimated to reduce the emittance to a well-defined value within the acceptance of the beamline and to establish the 1:1 imaging between the collimators, ESS and the coupling point.

In the PROSCAN facility, the emittance measurements are performed using the quadrupole scan method \cite{Sheppard1983, Schippers2002}.\ For a subset of energies, the emittance was reconstructed varying the magnetic field of quadrupole QD6 and recording the corresponding beam size changes with the monitors MD5X and MD6Y (\figref{fig:Envelope2}). 

For the transport line toward Gantry 3, a Mylar foil of 0.75 mm thickness was installed downstream of the ICC (see \figref{fig:Envelope1}).\ This foil mimics the scattering effect on the beam of a secondary electron current monitor that should be installed there.\ \tabref{table:emit} summarizes the results of the beam emittance measurements and the OPAL simulations with and without the effect of Mylar foil. 

\begin{table}[h!]
\caption{Comparison between measured and simulated transverse un-normalized total emittances \\ ($\pi$ mm mrad) with and without the Mylar foil}
\begin{ruledtabular}
 	\begin{tabular}{lcc|cr}
		 & \multicolumn{4}{c}{\textbf{Horizontal Plane}} \\
		 & \multicolumn{2}{c|}{\textbf{With Mylar foil}} & \multicolumn{2}{c}{\textbf{Without Mylar foil}}\\
		 \hline
		 \hline
		\textbf{Energy} & \textbf{Measur.} & \textbf{OPAL} & \textbf{Measur.} & \textbf{OPAL}\\	
		230 MeV	& 17.26	$\pm$ 0.35 &	18.02	& 16.65	$\pm$ 0.33 &	17.57 \\
		190 MeV & 19.64	$\pm$ 0.39 &	21.82	& 19.43	$\pm$ 0.39	& 20.88\\
		150 MeV & 27.19	$\pm$ 0.54 &	28.50	& 26.52	$\pm$ 0.53	& 27.36 \\
		110 MeV & 35.01	$\pm$ 0.71 &	37.03	& 34.53	$\pm$ 0.69	& 36.48 \\
		70 MeV  & 34.14	$\pm$ 0.68 &	39.95	& 32.23	$\pm$ 0.64	& 39.41 \\
		\hline
		\hline
		& \multicolumn{4}{c}{\textbf{Vertical Plane}} \\
		 & \multicolumn{2}{c|}{\textbf{With Mylar foil}} & \multicolumn{2}{c}{\textbf{Without Mylar foil}}\\
		 \hline
		 \hline
		\textbf{Energy} & \textbf{Measur.} & \textbf{OPAL} & \textbf{Measur.} & \textbf{OPAL}\\
		230 MeV	& 19.49	$\pm$ 0.39 &	19.40	& 19.01	$\pm$ 0.38 &	18.73 \\
		190 MeV	& 25.03	$\pm$ 0.50 &	23.42	& 24.63 $\pm$ 0.49 &	22.97\\
		150 MeV	& 27.95	$\pm$ 0.56 &	27.36	& 27.72	$\pm$ 0.55 &	26.85 \\
		110 MeV	& 32.74	$\pm$ 0.65 &	32.25	& 32.45	$\pm$ 0.65 &	31.92 \\
		70 MeV	& 36.25	$\pm$ 0.73 &    35.59	& 35.90	$\pm$ 0.72 &	34.63 \\
	\end{tabular}
\label{table:emit}
\end{ruledtabular}
\end{table}

As expected, a slightly higher beam emittance due to the Mylar foil was measured.\ The same effect was also reconstructed with the model, proving the efficacy of the OPAL particle-matter interaction model also with a thin foil.\ The difference between the measurements and the OPAL calculations increases for lower energies.\ However, averaging over the energy range, an increase of 2.1\% in the transversal emittance from the measurements was found, in good agreement with 2.4\% found in simulation.

This analysis constitutes another example of the importance to include particle-matter interaction into the beam dynamics model.


\subsection{Transmission}
\label{sec:trans}

In a proton therapy facility, the correct evaluation of the beamline transmission is of great importance since it is related to the dose delivered to the patient.\ As explained in section \ref{sec:2}, the transmission through the beamline varies almost by a factor of $10^3$ in the energy range 230$-$70 MeV.\ To treat the patient with the same beam intensity (typically 0.5$-$1 nA), the intensity compensation scheme was developed.\

In the beamline toward Gantry 3, the beam current and, hence, the transmission can be measured using the ionization chambers or evaluated from the profile strip monitors.\ In the second case, a calibration factor has to be applied to convert the signal from the monitor into the corresponding beam current ($I_{\text{beam}}$) \cite{Doelling2012}.\ Knowing the initial beam current from the cyclotron Comet ($I_{\text{Comet}}$) and calculating $I_{\text{beam}}$ from the last profile monitors before the coupling point of Gantry 3, the total transmission $T$ was evaluated as

\begin{equation}
 T = \frac{I_{\text{beam}}}{I_{\text{Comet}}}
\end{equation}

\begin{figure}[h]
\centering
\includegraphics[scale=0.17, keepaspectratio=true]{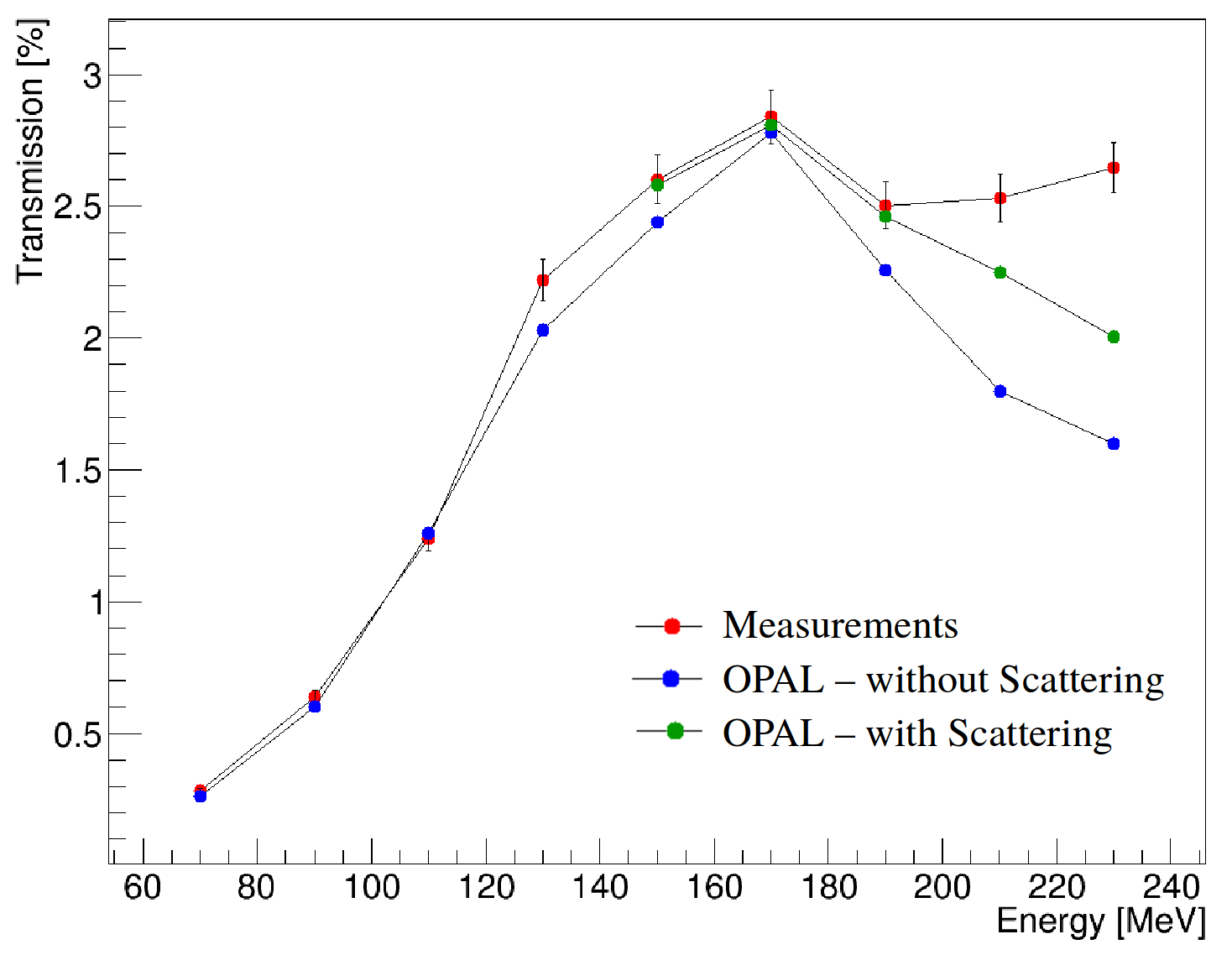}
\caption{Total transmission at the end of the beamline toward Gantry 3: comparison between the measurements and the OPAL model with and without the particle-matter interaction applied on the collimators.}
\label{fig:TransMess}
\end{figure}

The measured transmission was compared with the results from the OPAL model, as shown in \figref{fig:TransMess}.\ The trend of the transmission curve reveals the effect of the intensity compensation: an almost constant transmission is achieved above 140 MeV, while it decreases below 140 MeV.

An agreement better than 10\% was found between the OPAL model and the measurements in the low energy range, up to 150 MeV.\ In the high energy range, the discrepancy increases up to 20\% due to the scattered particles from the collimators along the beamline (blue points in \figref{fig:TransMess}).\ If this effect is not included in the optics model, it leads to an improper evaluation of the transmission, especially at higher energies where the scattering contribution is relevant.\ 

In the higher energy range, the effect of the scattering was included in the model by enabling the particle-matter interaction model to the collimators in the beamline.\ As shown in \figref{fig:TransMess} (green points), the total transmission from the OPAL model increases, reducing to 10\% the discrepancy with the measured value.\ This result reveals the importance of including scattering in addition to the energy loss in the beam dynamics model and encourages the further development of the particle-matter interaction model in OPAL; in particular, on the implication that the time integration has on the simplified elastic scattering model implemented.


\section{Conclusion}

The beam dynamics studies in proton therapy are normally dedicated to the development of the beam dynamics models for the accelerator \cite{Baumgarten2007, Benedetti2017} or for the dose delivery system (i.e.\ the last part of the beamline before the patient) \cite{Paganetti2004, Paganetti2011}.\ In this work, we develop an accurate beam dynamics model including the particle-matter interaction for the transport line that connects the accelerator with the dose delivery system.\\ 
In the preliminary design of a new beam transport line, TRANSPORT is a suitable and convenient beam dynamics code.\ However, in a proton therapy facility, the transmission, emittance and energy changes along the beamline are of primary importance.\ For this reason, the multi-particle beam dynamics code OPAL has been used for the first time to develop a more accurate and complete model of a proton therapy beamline.\ In addition, a direct comparison between the simulated particle distribution and the measured beam profile is also possible, allowing a better understanding of the beamline behaviour, as discussed in section \ref{subsec:ScatOnCol}.\\
The comparison between the model and the different types of measurements proves that the OPAL features are suitable to provide an accurate model of a transport line.\ The extension of the model to particle-matter interaction effects reduces the gap between the theoretical model and the measurements and allows for a better understanding of the beamline behaviour, as shown in section \ref{sec:4}.\ With TURTLE, similar simulations can be done, however the distributions after scattering and energy loss must be derived separately and included in each run and at each relevant location along the beamline.\ In this context, the use of a single code that combines Monte Carlo simulations with particle tracking is preferred and more convenient.\\
In the context of the intensity compensation, the measured transmission seems to be promising.\ After additional measurements and validations, this beam dynamics strategy would then be applied also to other transport lines of PROSCAN such as toward Gantry 2 or Optis 2.\ To develop this new beam optics the multi-objects optimizer available in OPAL can be used \cite{opt-pilot}.\ Starting from a complete beam dynamics model that considers also the particle-matter interaction, the solutions of the optimiser will reflect a more realistic situation in the beamline.\\
ROGER contributed to simplify the development of the model and the data analysis.\ The connection with the database and control system of PROSCAN provides a direct access to the beamline settings and measured beam profiles.\ With the post-processing tools, a full beam dynamics analysis can be performed on the model and the results compared with beam profile measurements.\\ 
In summary, a precise and complete beam dynamics model has been developed for the beam line that connects Comet to the coupling point of the Gantry 3.\ The OPAL framework seamlessly combines the particle tracking and the interaction with matter, leading to a more realistic model.\ The model benchmark against energy, emittance and profile measurements shows that in all cases a remarkable gain in the model completeness and accuracy has been achieved.\ We expect that the use of a single tool like OPAL will simplify the detailed design steps of beam transport lines, normally done by combining several beam dynamics codes, each with its own strength and limitations. 


\begin{acknowledgments}
The authors wish to thank H. Lutz for the improvement of the PROSCAN database and the connection with the OPAL model.\ We also thank R. D\"{o}lling for many useful discussions and valuable comments concerning this work and T. Schietinger for the proofreading of the manuscript. 
\end{acknowledgments}

\nocite{*}

\bibliography{biblio}

\begin{thebibliography}{10}

\bibitem{VanGoethem2009}
M.J. van Goethem~\textit{et al.}
\newblock {Geant4 simulations of proton beam transport through a carbon or
  beryllium degrader and following a beam line}.
\newblock {\em Phys. Med. Biol.}, 54(19):5831--46, 2009.

\bibitem{Pavlovic2008}
M.~Pavlovic and I.~Strasik.
\newblock {Beam Transport with Scattering using SRIM supporting Software
  Routines Code}.
\newblock Number TUPP102, 2008.
\newblock EPAC08.

\bibitem{Farley2005}
J.~M. Farley.
\newblock {Optimum strategy for energy degraders and ionization cooling}.
\newblock {\em Nucl.Inst.Meth. A}, 540(2-3):235--244, 2005.

\bibitem{Anferov2003}
V.~Anferov.
\newblock {Energy degrader optimization for medical beam lines}.
\newblock {\em Nucl. Instr and Meth. A}, 496:222--227, 2003.

\bibitem{Paganetti2011}
H.~Paganetti.
\newblock {\em {Proton Therapy Physics}}.
\newblock {CRC Press, Taylor \& Francis Group}, 2011.

\bibitem{Transport}
U.~Rohrer.
\newblock {\em {PSI Graphic Transport Framework}}.
\newblock based on a CERN-SLAC-FERMILAB version by K.L. Brown \textit{et al.}

\bibitem{Turtle}
U.~Rohrer.
\newblock {\em PSI Graphic Turtle Framework}.
\newblock based on a CERN-SLAC-FERMILAB version by K.L. Brown \textit{et al.}

\bibitem{Madx}
http://mad.web.cern.ch/mad/.

\bibitem{Geant}
S.~Agostinelli \textit{et al.}
\newblock {Geant4 - A simulation toolkit}.
\newblock {\em Nucl. Instr and Meth. A}, 506:250--303, 2003.

\bibitem{Fluka}
A.~Ferrari, P.R. Sala, A.~Fass{\'o}, and J.~Ranft.
\newblock {\em {FLUKA: a multi-particle transport code}}, 2005.
\newblock CERN Yellow report, CERN-2005-10.

\bibitem{Topas}
http://www.topasmc.org/.

\bibitem{Paganetti2004}
H.~Paganetti, H.~Jiang, S.Y. Lee, and H.M. Kooy.
\newblock {Accurate Monte Carlo simulations for nozzle design, commissioning
  and quality assurance for a proton radiation therapy facility}.
\newblock {\em Med. Phys.}, 31(7):2107--18, 2004.

\bibitem{Rohrer2002}
U.~Rohrer.
\newblock {Optical design of the proton beam lines for the PROSCAN project}.
\newblock Technical Report Volume V, Paul Scherrer Institut, 2002.
\newblock http://aea.web.psi.ch/Urs\textunderscore
  Rohrer/MyWeb/pdfs/proscan.pdf.

\bibitem{OPAL}
A.~Adelmann, A.~Gsell (PSI), C.~Metzger-Kraus (HZB), Y.~Ineichen (IBM),
  S.~Russel (LANL), C.~Wang, J.~Yang (CIAE), S.~Sheehy, C.~Rogers (RAL), and
  D.~Winklehner (MIT).
\newblock {The OPAL (Object Oriented Parallel Accelerator Library) Framework}.
\newblock Technical Report PSI-PR-08-02, Paul Scherrer Institut, 2008 - 2017.

\bibitem{H5root}
T.~Schietinger.
\newblock {H5PartROOT - A visualization and post-processing tool for
  accelerator simulations}.
\newblock Number {THPSC049}, pages (343--346), 2009.

\bibitem{Pedroni1995}
E.~Pedroni \textit{et al.}
\newblock {The 200-MeV proton therapy project at the Paul Scherrer Institute:
  Conceptual design and practical realization}.
\newblock {\em Med. Phys.}, 2(1):37--53, 1995.

\bibitem{Pedroni2004}
E.~Pedroni \textit{et al.}
\newblock {The PSI Gantry 2: a second generation of proton scanning Gantry}.
\newblock {\em Z. Med. Phys.}, 14(1):25--34, 2004.

\bibitem{Pedroni2011}
E.~Pedroni, D.~Meer, C.~Bula, S.~Safai, and S.~Zenklusen.
\newblock {Pencil beam characteristics of the next-generation proton scanning
  gantry of PSI: design issues and initial commissioning results}.
\newblock {\em Eur. Phys. J. Plus}, 126:66--93, 2011.

\bibitem{optis}
https://www.psi.ch/protontherapy/optis-2.

\bibitem{Varian}
http://www.varian.com.

\bibitem{PIF}
http://pif.web.psi.ch.

\bibitem{Koschik2015}
A.~Koschik \textit{et al.}
\newblock {Gantry 3: Further development of the PSI PROSCAN Proton Therapy
  Facility}.
\newblock Number ISBN 978-3-95450-168-7, pages (2275--2277), 2015.
\newblock IPAC15.

\bibitem{Schippers2007}
J.M.~Schippers \textit{et al.}
\newblock {The SC cyclotron and beam lines of PSI's new protontherapy facility
  PROSCAN}.
\newblock {\em Nucl. Instr and Meth. B}, 261(1):773--776, 2007.

\bibitem{Baumgarten2007}
C.~Baumgarten \textit{et al.}
\newblock {Isochronism of the ACCEL 250 MeV medical proton cyclotron}.
\newblock {\em Nucl. Instr and Meth. A}, 570(1):10--14, 2007.

\bibitem{Baumgarten2015-2}
C.~Baumgarten.
\newblock {Gantry 3 Transmission}.
\newblock Technical report, Paul Scherrer Institut, 2015.

\bibitem{Schippers2007-1}
J.M.~Schippers \textit{et al.}
\newblock {First year of operation of PSI's new cyclotron and beam lines for
  proton therapy}.
\newblock Number MOXCR04, 2007.
\newblock Cyclotron Conference Catania.

\bibitem{Baumgarten2015}
C.~Baumgarten and A.~Gerbershagen.
\newblock {Gantry 3 Beamline Commissioning up to UPG3}.
\newblock Technical Report BC85-0010a, Paul Scherrer Institut, 2015.

\bibitem{Stachel2013}
H.~Stachel.
\newblock {Double Degrader for Proton Therapy}.
\newblock Master's thesis, ETH Z{\"u}rich, 2013.
\newblock
  http://amas.web.psi.ch/people/aadelmann/ETH-Accel-Lecture-1/projectscompleted/phys/stachel.pdf.

\bibitem{Locans2016}
A.~Adelmann, U.~Locans, and A.~Suter.
\newblock {The Dynamic Kernel Scheduler-Part 1}.
\newblock {\em Computer Physics Communications}, 207:83--90, 2016.

\bibitem{Enge}
H.~A. Enge.
\newblock {\em {\textit{Deflecting Magnets} in Focusing of Charged Particles}}.
\newblock {A. Septier, Academic Press}, 1967.

\bibitem{Leo}
W.R. Leo.
\newblock {\em Techniques For Nuclear And Particle Physics Experiments}.
\newblock Springer-Verlag, second revised edition edition, 1994.

\bibitem{ICRU}
ICRU Bethesda.
\newblock {Stopping Powers and Ranges for Protons and Alpha Particles}, 1993.
\newblock ICRU-49.

\bibitem{Andersen}
H.~H. Andersen and J.~F. Ziegler.
\newblock {\em {Hydrogen Stopping Power and Ranges in all Elements}}.
\newblock Pergamon, New York, 1977.

\bibitem{PDG}
K.A.~Olive \textit{et al.}
\newblock {\em Particle Data Group}.
\newblock Chin. Phys. C, 2014.

\bibitem{Jackson}
J.D. Jackson.
\newblock {\em Classical Electrodynamics}.
\newblock John Wiley and Sons Ltd., third edition edition, 1962.

\bibitem{Reist2002}
H.~Reist \textit{et al.}
\newblock {A fast degrader to set the energies for the application of the depth
  dose in proton therapy}.
\newblock Paul Scherrer Institut, Annual Report, Volume V, 2002.

\bibitem{Doelling2015}
R.~D{\"o}lling.
\newblock {Energie nach dem Degrader}.
\newblock Technical Report P24/DR84-1505.0, Paul Scherrer Institut, 2015.

\bibitem{Doelling2012}
R.~D{\"o}lling.
\newblock {Faktoren zur Berechnung des passierenden Strahlstroms aus den
  Signalstr{\"o}men von d{\"u}nnen Strommonitoren, Standard-Profilmonitoren,
  Stoppern, Halo- und Schwerpunktmonitoren}.
\newblock Technical Report P24/DR84-620.5, Paul Scherrer Institut, 2012.

\bibitem{VanLuijk2001}
H.D.~Zelle P.~van Luijk, A. A. van\'t~Veld and J.M. Schippers.
\newblock {Collimator scatter and 2D dosimetry in small proton beams}.
\newblock {\em Phys. Med. Biol}, 46(3):(653--670), 2001.

\bibitem{Sheppard1983}
J.C.~Sheppard \textit{et al.}
\newblock {Emittance calculations for the Stanford Linear Collider injector}.
\newblock {\em IEEE Trans. Nucl. Sci.}, NS-30(4), 1983.

\bibitem{Schippers2002}
J.~M. Schippers.
\newblock {Accuracy assessment of some methods for emittance measurement}.
\newblock Technical report, Paul Scherrer Institut, 2002.

\bibitem{Benedetti2017}
S.~Benedetti, A.~Grudiev, and A.~Latina.
\newblock High gradient linac for proton therapy.
\newblock {\em Phys. Rev. Accel. Beams}, 20:040101, 2017.

\bibitem{opt-pilot}
Y.~Ineichen.
\newblock {\em Toward Massively Parallel Multi-objective Optimization with
  Application to Particle Accelerators}.
\newblock PhD thesis, ETH Z{\"u}rich, 2013.
\newblock (21114).

\end{thebibliography}
\bibliographystyle{unsrt}

\end{document}